\documentclass[vecphys]{svmult}
\usepackage{ergebnisse} 

\usepackage{graphicx}        
\usepackage{multicol}        
\usepackage[bottom]{footmisc}

\usepackage[authoryear]{natbib} 
\usepackage{revsymb} 

\def\gtrsim{{_>\atop^{\sim}}}
\def\zbar{\rlap{$z$}{\raise 1.0ex\hbox{-}}}
\newcommand\aj{AJ}
\newcommand\araa{ARA\&A}
\newcommand\apj{ApJ}
\newcommand\apjl{ApJL}
\newcommand\apjs{ApJS}
\newcommand\apss{Ap\&SS}
\newcommand\aap{A\&A}
\newcommand\aaps{A\&AS}
\newcommand\mnras{MNRAS}
\newcommand\pasp{PASP}
\newcommand\qjras{QJRAS}
\newcommand\nat{Nature}
\newcommand\physrep{Phys.~Rep.}

{

\begin{document}

\setcounter{tocdepth}{2}
\setcounter{secnumdepth}{3}

\renewcommand{\chaptermark}[1]{\markboth{\chaptername%
\ \thechapter:\,\ #1}{}}
\renewcommand{\sectionmark}[1]{\markright{\thesection\,\ #1}}

\title*{Ultraluminous Infrared Galaxies}

\author{Carol J. Lonsdale, Duncan Farrah \& Harding E. Smith}

\institute{} 
       
\authorrunning{Lonsdale, Farrah \& Smith}

\maketitle
\begin{abstract}
Ever since their discovery in the 1970's, UltraLuminous InfraRed
Galaxies (ULIRGs; classically $L_{ir} > 10^{12}L_{\odot}$) have
fascinated astronomers with their immense luminosities, and
frustrated them due to their singularly opaque nature, almost in
equal measure. Over the last decade, however, comprehensive
observations from the X-ray through to the radio have produced a
consensus picture of local ULIRGs, showing that they are mergers
between gas rich galaxies, where the interaction
triggers some combination of dust-enshrouded starburst and AGN
activity, with the starburst usually dominating. Very recent
results have thrown ULIRGs even further to the fore. Originally
they were thought of as little more than a local oddity, but the
latest IR surveys have shown that ULIRGs are vastly more numerous
at high redshift, and tantalizing suggestions of physical differences
between high and low redshift ULIRGs hint at differences in their
formation modes and local environment. In this review we look at
recent progress on understanding the physics and evolution of
local ULIRGs, the contribution of high redshift ULIRGs to the
cosmic infrared background and the global history of star
formation, and the role of ULIRGs as diagnostics of the formation
of massive galaxies and large-scale structures.
\end{abstract}

\section{Introduction}
 
\subsection{The Biggest and the Brightest}

In human endeavour there's a fascination with the
``biggest and the best'', or the ``best and the brightest''. It's a matter
for the social and psychological sciences to speculate on the reasons we
feel driven to give Oscars to the best movies and to climb the highest
mountains, but being human (well, most of them anyway), astrophysicists are
not immune to the desire to search for the Universe's own brand of the
biggest and the brightest. We tend to give them extreme names such as {\it
ultra}-this or {\it hyper}-that. In their rarity however, unusual objects
and environments teach us about the most extreme physical processes in
the Universe.
 
Which brings us to ULIRGs (or ULIGs to some) - Ultra Luminous InfraRed
Galaxies.  There are also HLIRGs (or HiLIRGs or HyLIRGs, or indeed HyLIGs):
Hyper Luminous InfraRed Galaxies.  These Oscar contenders have historically
been defined simply in terms of luminosity: $L_{8-1000{\mu}m}$ between
10$^{12}$ and 10$^{13} L_{\odot}$ for the ULIRGs and $> 10^{13}$
$L_{\odot}$ for HLIRGs \footnote{The supporting cast
consists of LIRGs (or LIGs), the much more common lower luminosity
understudies of the prima donnas, with $L_{8-1000{\mu}m}$ between $10^{11}$
and $10^{12} L_{\odot}$}.  What are these dramatic objects, and why are
they among the brightest objects in the Universe? Does this also imply they
are the biggest in term of mass or size, or are they just superficial
fireworks that leave little lasting impression? What can understanding
these enigmatic objects teach us about the evolution of galaxies and how
the the Universe came to look as it now does?
 
The last few years have seen a dramatic shift in our perceptions of
ULIRGs. Once believed by many to be a rare curiosity -- certainly
interesting, yes, but perhaps no more than a local oddity -- they're
beginning to see center stage much more frequently.  Ironically, this
increased interest in these rare objects has arisen because we now realize
that ULIRGs were once not nearly so rare as we find them to be in the local
Universe: pioneering submillimeter and millimeter surveys have demonstrated
that ULIRGs are many hundreds of times more numerous at z$>$1 than they are
locally.  This in turn suggests that they played a much more important role
in galaxy formation and evolution than we imagined, and so understanding
them becomes of prime importance. Fortunately, this rise in the fortunes of
the ULIRG has occurred in an era when many new observing capabilities are
coming on line.  Foremost among these is the Spitzer Space Telescope, with
its suite of deep imaging infrared cameras and its sensitive infrared
spectrograph. Spitzer's extensive first results on ULIRGs are just now
beginning to be published. Submillimeter and millimeter cameras are also
improving dramatically; upcoming facilities include AzTEC, SCUBA-2,
Herschel, and ALMA. We can also look forward to major new insights from
mid- and near-IR facilities such as ASTRO-F, WISE, and JWST, as well as
radio facilities such as the Square Kilometer Array, and new X-ray
facilities with high hard X-ray sensitivities such as Con-X and XEUS.  The
timeliness of this subject is exemplified by recent reviews of The Cosmic
InfraRed Background \citep{lpd05}, Interacting Galaxies \citep{vang05},
Megamasers \citep{lo05}, Galactic Winds \citep{vei4}, and High Redshift
Molecular Gas \citep{vandenb05}, all of relevance to ULIRGs.

\subsection{Overview} 

ULIRGs were first discovered in large numbers by the Infrared Astronomical
Satellite in 1983, and were found to be comparatively rare locally, with a
space density several orders of magnitude lower than that of normal
galaxies, and possibly a factor of a few higher than QSOs. Followup
observations show that most, if not all ULIRGs are found in major disk
mergers, and that the central few hundred pc of their nuclear regions
harbour very large masses of gas and dust. The power source behind the IR
emission is some combination of a large population of hot young stars (a
`starburst'\footnote{for our purposes defined as a star forming event with
a gas exhaustion timescale very short compared to the Hubble time}) or a
very massive black hole accreting matter at a rapid rate (which for the
remainder of this review we refer to as an `AGN').  Though distinguishing
between the two initially (and even now) proved to be very difficult, it is
now thought that, at least locally, ULIRGs are mainly powered by a
starburst, but frequently with a significant AGN contribution. Local ULIRGs
reside in relatively low-density environments (not
unexpectedly, since relative velocities are thought to be too high for
mergers to occur in rich, virialized environments), and are expected to
evolve into spheroidal systems as the galaxy mergers that appear to trigger
ULIRG activity progresses.

Even IRAS was sensitive enough to determine that there has been very strong
evolution in the ULIRG (and LIRG) population with redshift out to at least
$z\sim0.5$, with approximate form $(1+z)^4$.  IRAS also found ULIRGs out to
extremely high redshifts, including the famous, lensed, F10214$+$4724 at
z=2.286.  This strong evolution was confirmed with results from the
Infrared Space Observatory which, although covering much smaller areas than
IRAS, could probe this evolution out to $z\sim1$ due to its greater
sensitivity (Figure 1).  This evolution was later recast as the now
ubiquitous `star formation history of the Universe' figures, which show
that LIRGs rather than ULIRGs are responsible for the bulk of the evolution
seen since z$\sim$1 in IR galaxies. ULIRGs, however, did not slink into the
shadows; on the contrary they returned triumphant with the advent of
submillimeter imaging surveys, which came shortly after ISO and can in
principle probe IR-luminous galaxies up to $z\sim7$. These sub-mm surveys
showed that ULIRGs are orders of magnitude more numerous at $z>1$ than
locally, outnumbering optically bright QSOs at those redshifts by a large
margin. Followup observations showed that these distant ULIRGs bear many
similarities to their local cousins, but also exhibit some key differences,
and that they may signpost the obscured phases of the very dramatic events
suspected of building the most massive galaxies seen in the local Universe.

When considered within the framework of modern theories for the formation
of galaxies and large-scale structure, it seems initially surprising that
there are many more ULIRGs at high redshift than locally, because in early
implementations of the `hierarchical buildup' paradigm, large galaxies
build up slowly from the mergers of smaller systems.  This is in contrast
to the early `monolithic collapse' models \citep{egg62}, where ellipticals
formed early in a dramatic burst of star formation, which had been largely
supplanted in favour of hierarchical models.  The discovery of so many
ULIRGs at high redshifts caused hierarchical models of the time major
difficulties in making enough distant systems with such high star formation
rates. The basic dark matter halo growth theory, described by an
extended/modified Press-Schechter formalism, does however allow for rapid
baryon accumulation in very massive dark matter halos, and recent galaxy
formation models are having greater success in producing the observed
number of ULIRGs in sub-mm surveys, albeit with some stringent assumptions.

In this review, we will therefore focus on selected key topics: (1) our
understanding of the astrophysics of local ULIRGs, and in particular the
relative importance of star formation versus AGN in powering ULIRGs, (2)
similarities and differences between local and high-redshift ULIRGs, and
(3) the relationship between ULIRGs and the
formation of large-scale structure and of galaxies as a function of
redshift.  Since the study of
ULIRGs clearly connects to many major disciplines of observational and
theoretical extragalactic astronomy we cannot hope to cover all topics of
relevance to them in this review.  Nor can we completely review all recent
published studies of ULIRGs; excellent ULIRG papers simply abound.  We
therefore highlight the most recent advances and our perspective on the
most important questions concerning their study within the framework of
galaxy and structure formation.

In section \ref{sec:discover} we provide brief historical context to the
discovery of ULIRGs and their evolutionary role. In section
\ref{sec:physics} we review current understanding of the astrophysics of
local ULIRGs by wavelength, and in section \ref{sec:picture} we summarize
local studies into a picture of ULIRG nature and evolution in the local
Universe.  In section \ref{ulirghigh} we review observations of ULIRGs at
higher redshifts, based primarily on data from ISO, SCUBA, HST and
Spitzer. Section \ref{ulirglss} places these studies into the context of
structure formation and reviews their role within galaxy formation
scenarios. Finally, section \ref{sec:future} highlights the key open
questions and our perspectives of where the answers are likely to come
from.
 
For the remainder of this review we usually refer to ULIRGs and HLIRGs
combined as ULIRGs, because many of the earlier works used the term
``ULIRG'' to refer to all objects above 10$^{12}$ in $L_{\odot}$ and the
term HLIRG has been in only recent and inconsistent use. We assume
$H_{0}=70$ km s$^{-1}$ Mpc$^{-1}$, $\Omega=1$, and
$\Omega_{\Lambda}=0.7$. Luminosities are quoted in units of bolometric
solar luminosities, where $L_{\odot} = 3.826\times10^{26}$ Watts. Unless
otherwise stated, the term `IR' or `infrared' luminosity refers to the
integrated rest-frame luminosity over 1-1000 or 8-1000$\mu$m (which differ
very little for most SEDs).

\begin{figure}
\centering
\includegraphics[height=7cm]{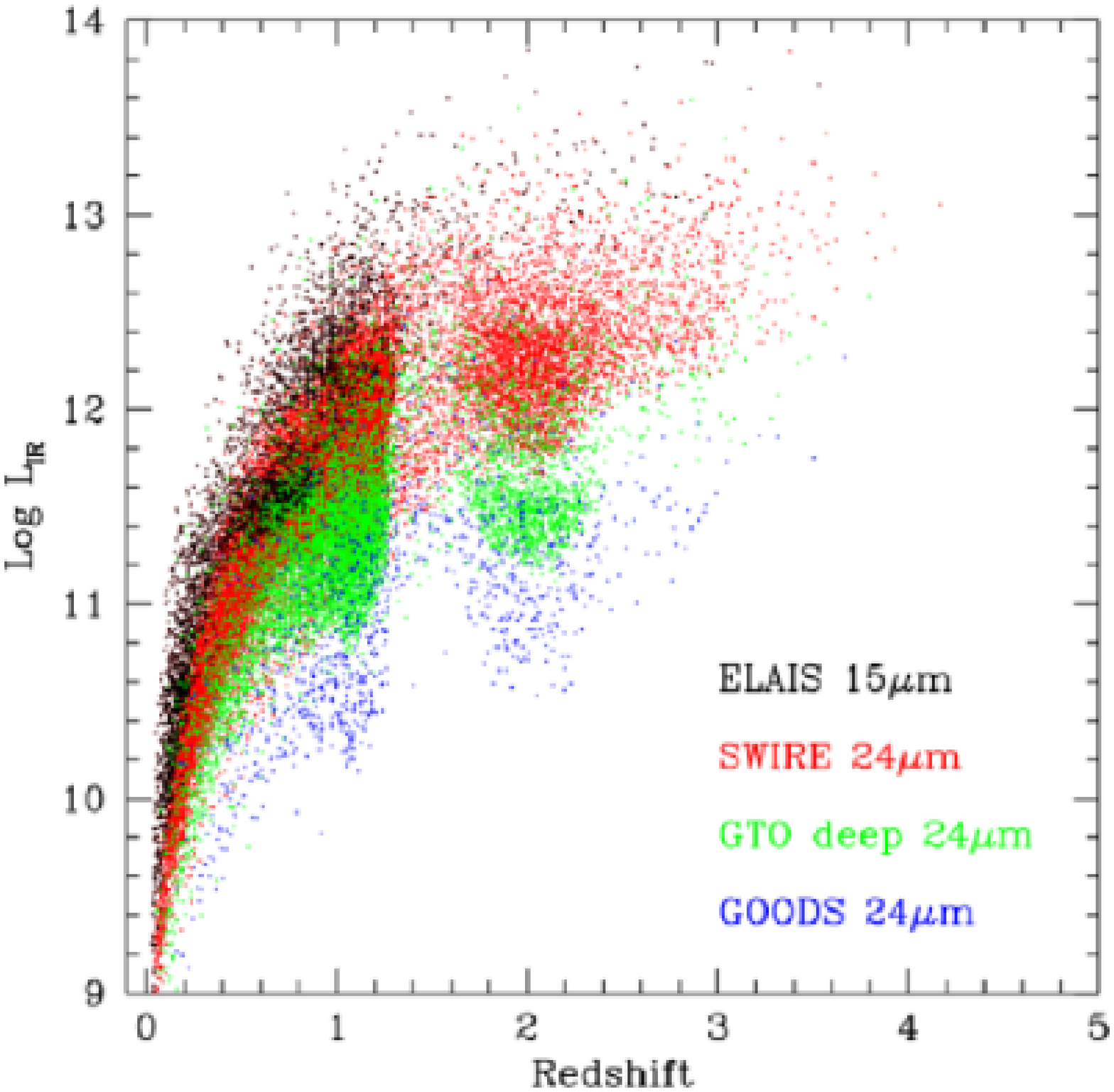}
\includegraphics[height=7cm]{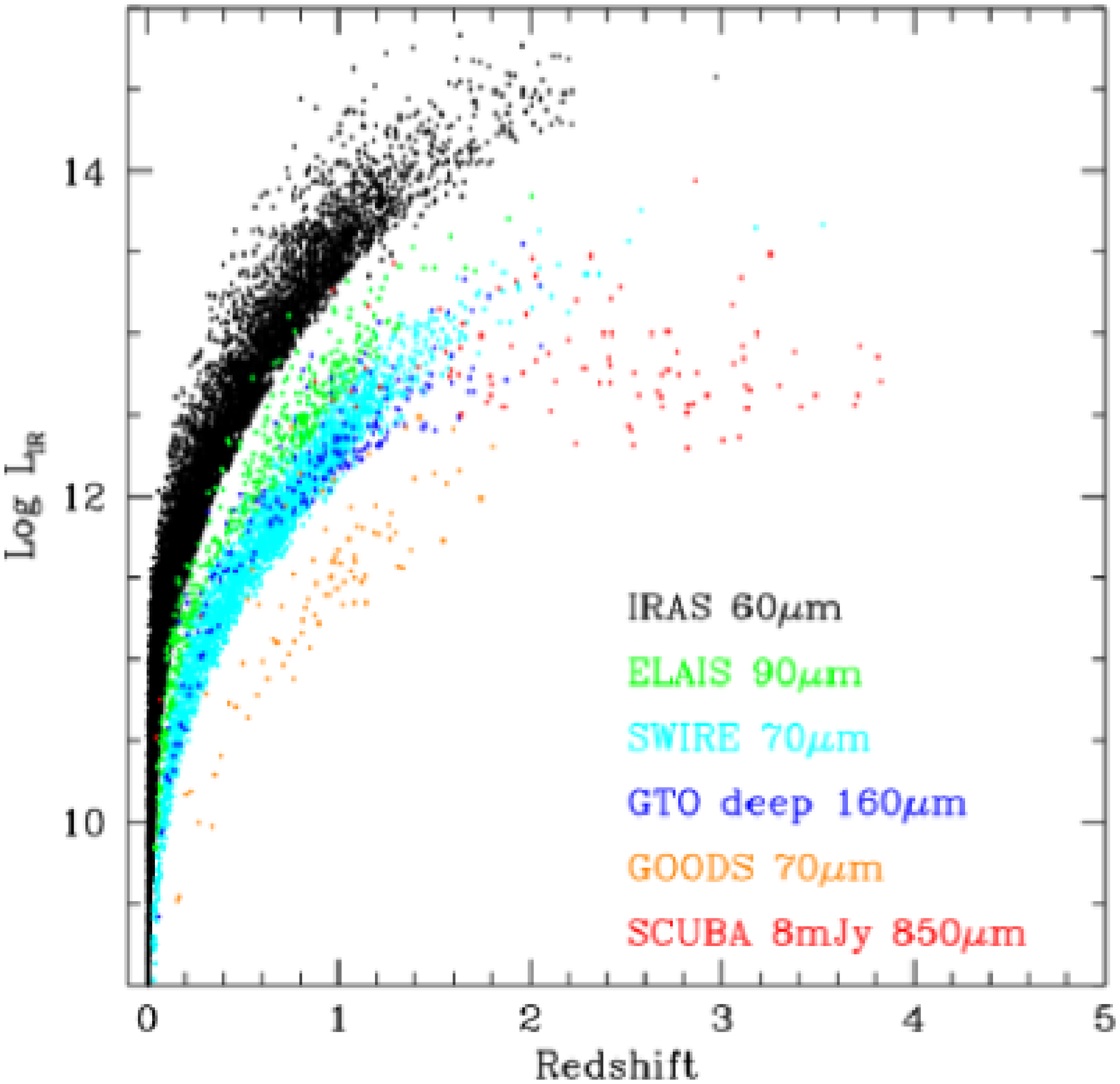}
\caption{Simulation of the ability of recent infrared surveys to discover ULIRGs, with
shorter wavelength ($\lambda < 25\mu$m) surveys plotted in the upper panel
and longer wavelength surveys below. The band in the upper
figure at $z \sim 1.5$ is produced by the the 10$\mu$m silicate
absorption feature which falls in the Spitzer 24$\mu$m band at that 
redshift.  Wide 
shallow surveys have the largest volume for discovering the most luminous
ULIRGs, while
narrow deep surveys can of course find the most distant ones, though in
much smaller number. Tiered ('wedding cake') surveys are thus required
to construct complete luminosity functions within a given redshift interval
(vertical slices). Based on the simulations of \citet{xu03}, these
simulations can fit IRAS, ISO, Spitzer and submillimeter counts and
redshift distributions. We plot 1/5 of the objects expected within the area
and depth of each survey: ISO ELAIS 15$\mu$m \citep{vac}; SWIRE 24 \& 70$\mu$m
\citep{sur05}; Spitzer Guaranteed Time Observer deep 24$\mu$m
\citep{perz}; GOODS 24$\mu$m \citep{char04}; IRAS 60$\mu$m \citep{lon90};
SCUBA 850$\mu$m (Scott et al. 2002); GOODS 70$\mu$m (D. Frayer, priv.
comm.); Spitzer GTO deep 160$\mu$m \citep{dol2}.}

\label{lzplots}       
\end{figure}

\section{The discovery of ULIRGs}\label{sec:discover}
It is often stated that UltraLuminous InfraRed Galaxies were discovered by
the InfraRed Astronomy Satellite (IRAS), which was launched in
1983. Strictly speaking this may be true since the first objects to meet
the now generally acknowledged criteria which define a ULIRG -
$L_{8-1000{\mu}m}>10^{12}$ L$_{\odot}$ and, optionally, exceptionally large
ratios of $L_{infrared}/L_{optical}$ - can indeed be considered to have
been discovered with the publication by \citet{hou85} of a sample of
9 IRAS sources invisible or exceptionally faint on the Palomar Sky Survey
plates, and exhibiting $L_{ir}/L_{opt}$ ratios over 50. We must go back
further, however, to appreciate the origins of the ULIRG
phenomenon. Galaxies with an unusually high amount of mid- or far-infrared
emission compared to their optical output had been discovered by the
pioneers of infrared astronomy in the 10 to 15 years preceding IRAS, and
had been recognised to represent luminous, probably short-lived, active
events of some sort within a galaxy.
 
Several excellent reviews have described the IRAS and pre-IRAS legacy,
including Rieke and Lebofsky (1979), \citet{soi87}, Sanders \& Mirabel (1996), 
and \citet{mir02} so we present here only the brief highlights relevant to 
ULIRG discovery and ULIRG samples.
 
The first infrared observations of galaxies were undertaken in the late
sixties \citep{lk68,kl70}, and a population of galaxies with
infrared-dominant emission from the nuclear regions was discussed by
\citet{rie}.  The well known radio-infared relation \citep{dej85,hel} 
was originally suggested by van der Kruit (1971).  An early debate ensued
as to the thermal vs non-thermal origin of the infrared emission from
galaxies \citep{rees,bs70} with the conclusion that most systems are
powered by re-radiation of starlight by dust (Rieke \& Lebofsky 1979 and
references therein). The other key contribution of the pre-IRAS days was
the role of interactions in triggering nuclear and starburst activity
\citep{too,lt78}, which was confirmed for the strongly interacting system
Arp 299 by \citet{gehrz}, and several other interacting and merging systems
by \citet{lon84,jow}.
 
IRAS scanned almost the entire sky in the thermal infrared
\footnote{defined as the wavelength region over which dust grains can
thermally re-radiate emission, ranging from the dust sublimation
temperature on the short wavelength side ($\sim$ 1$\mu$m) to $\sim$
1000$\mu$m on the long wavelength side where non-thermal processes often
begin to dominate}, observing in four bands centered at 12, 25, 60
and 100$\mu$m and opening up an unprecedented volume of space to study 
in this wavelength regime. This allowed IRAS to find
exceedingly rare objects and to demonstrate the importance of this new
class of exceptionally infrared-luminous objects (Figure 1). One of the
most exciting early discoveries was that IRAS would not be limited
to exploring only the very local Universe due to its relatively bright
flux limits, but it could reach out to significant distances thanks
to the existence of large numbers of exceptionally infrared-luminous
sources. Strong evolution was demonstrated for the IRAS deep
north ecliptic hole field by \citet{hac87} and confirmed over
the sky by \citet{lon90} and \citet{sau}. Extensive
redshift surveys \citep{oli96} led to the discovery of the first
z$>$2 IRAS HLIRG, IRAS FSC 10214+4724 at z=2.86 \citep{rr91}
which is a lensed system.  The extremely high evolution rates,
modeled, for example, as $L_{IR} {\sim} (1+z)^{{\sim}4}$, were actually
anticipated, based on the previously known strong evolution of
starburst-related sub-mJy radio sources \citep{hac87}.
 
ULIRGs were found to have comparable space densities to those of PG QSOs of
similar luminosity by \citet{soi87}; however, significant incompleteness in
the PG QSO sample has since been demonstrated \citep{wis00} so this result
needs re-visiting. The best known samples of IRAS luminous and
ultraluminous galaxies, all selected at 60$\mu$m, are the Bright Galaxy
Sample of \citet{soi87}, recently significantly updated into the Revised
Bright Galaxy Sample (RBGS; Sanders et al. 2005), and the complete
flux-limited IRAS 1 Jy sample \citep{ks98}.  Also notable are the 2Jy
sample of \citet{str} and the FIRST/IRAS sample of \citet{sta00}.  The RBGS
contains 629 IRAS 60$\mu$m galaxies brighter than 5.24 Jy and Galactic
latitude $>$5 degrees and contains 20 ULIRGs; the most luminous being Mrk
231 with $L_{ir} = 3.2\times10^{12}L_{\odot}$, the highest redshift being
IRAS 07251-0248 at z=0.0876, and the closest being Arp 220 at z=0.018. The
1 Jy sample consists of 118 ULIRGs drawn from the IRAS Faint Source
Catalog, with declination $>$-40 degrees and Galactic latitude $|b|>$30
degrees.  This sample also has a warm 60/100$\mu$m$>$0.3 colour selection,
which introduces a bias against cooler objects.
 
Whilst dramatic in nature, LIRGs and ULIRGs in the local Universe are 
rare, and they contribute only $\sim$6\% of the total infrared
luminous energy density \citep{sn91}, and about $\sim$3\% of the
total energy density.  It was found that both the IRAS 60/100$\mu$m colour
and the $L_{IR}/L_{opt}$ ratio increased with luminosity \citep{soi87},
reaching $\sim 100$ in the most extreme systems. This indicates
that the more luminous systems must have an increasing contribution from an
additional warm source compared to the relatively cool emission from modest
levels of star formation seen in spiral disks \citep{dej85}.  IRAS
ULIRGs have been separated into warm and cool subsamples based on the 25/60 and
60/100$\mu$m IRAS colours \citep{san3,sur}, with
the warmer objects more likely to host an AGN \citep{deg85}.
The spectral energy distributions of IRAS ULIRGS were reviewed by Sanders
and Mirabel (1996), illustrating these trends with colour and luminosity.
These reviewers also compared the SEDs of QSOs and Blazars with those of
ULIRGs.
 
It should be remembered that although the IRAS datasets are now nearly a
quarter of a century old, they are very under-explored.  As of publication
date there are $>$37,000 infrared-bright galaxies in the IRAS Faint Source
Catalog (FSC) that have never been observed with any other instrument and
reported in any journal article.  Only 43\% of the 64,606 IRAS extragalactic
FSC sources\footnote{As reported by NED, the NASA/IPAC Extragalactic
Database} have been included in any sort of publication (J. Mazzarella,
priv. comm.).

\section{The Physics of Local ULIRGs}\label{sec:physics}

The complexity of the ULIRG phenomenon requires a multi-wavelength approach
and we have learned an enormous amount from detailed studies at many
wavelengths.  For this review we focus on those observational areas
which we believe are key to understanding the ULIRG phenomenon, in part
because they represent the spectral regions of lowest optical depth, and in
which key advances have been made in the years since the Sanders \& Mirabel
review.  We stress, however, that even the relatively low-optical depth
radio, mid-infrared, and X-ray regimes may still show substantial
obscuration due to starburst-related free-free absorption \citep{con1}, 
compact nuclear molecular clouds or molecular tori, which may be
optically thick at wavelengths as long as 30$\mu$m, or Compton-thick
absorbing columns with $N_H \gtrsim 10^{24}$ cm$^{-2}$. 

The compactness of the
IR-emitting regions in luminous IR galaxies ({\it c.f.} Condon et al.
1991) suggests two possible origins for the high ULIRG luminosities:
compact nuclear starbursts and/or highly obscured AGN activity.  Much early
effort was given to determining which phenomenon powers ULIRGs; with the
understanding that starburst and AGN emission are frequently found together
in luminous galaxies, the emphasis has shifted to determining which
mechanism is dominant and to understanding the relationship between
co-existing AGN \& starburst emission.

\subsection{Optical to mid-IR Imaging}

Early imaging surveys of IR-luminous sources spanned a wide range in
luminosity (from $<10^{9}L_{\odot}$ up to $\sim10^{12}L_{\odot}$, and
revealed an interesting picture \citep{soi0,rie1}. Sources with
$L_{ir}<10^{9}L_{\odot}$ are almost exclusively confined to undisturbed E
and S0 systems, with few spirals. The fraction of spirals increases sharply
with increasing IR luminosity however, with most systems in the range
$10^{10}L_{\odot}<L_{ir}<10^{11}L_{\odot}$ being Sb or Sc type systems. At
luminosities above about $10^{11}L_{\odot}$, the majority of systems are
still spirals, but an increasing fraction (up to $\sim25\%$) appear to be
involved in interactions, or to show signs of morphological disturbance.
 
This apparent increase in the number of interacting IRAS systems was thrown
into sharp relief by early ULIRG imaging studies, showing that interactions
and mergers are much more common amongst ULIRGs than in lower luminosity
systems, though the exact fraction in ongoing interactions remained
contentious for some years. The first imaging surveys \citep{arm87} showed
that at least 70\% of systems with ULIRG or near-ULIRG luminosities are
interacting, with morphologies expected from the collision of two disk
galaxies. Later optical and near-IR studies of ULIRGs
\citep{mel90,hut91,cle96} found a higher fraction of ULIRGs involved in
interactions, at least 90\%, and that there are a wide range of merger
stages present in the ULIRG population, from widely separated systems to
advanced mergers \citep{mur96}. Other studies, however, found a much lower
fraction of ULIRGs involved in interactions, fewer than 70\%
\citep{law0,zou91,lee94}.
 
These results demonstrate that interactions and mergers play an important
role in the ULIRG population. This provided a plausible trigger for the
immense IR luminosities seen in ULIRGs. In order to both fuel and enshroud
the power sources in ULIRGs (irrespective of whether they are starbursts,
AGN, or both), a large quantity of gas and dust needs to be channeled into
a small volume, most plausibly sited in the nucleus of the host
galaxy. Results from N-body modeling of galaxy collisions (see Barnes \&
Hernquist 1992 for a review) suggest that this is readily achieved during
the course of a merger \citep{barn,bar96,mih1996,dub99}, though details
depend on many variables (e.g. angle of approach, relative velocity, disc
inclinations, bulge size and gas and dark matter masses). The majority of
mergers involve an initial close approach, followed by a maximum separation
of up to $\sim50$kpc, reached after $\sim2.5\times10^{8}$ years, then a
second close approach a few tens of millions of years later, which is
rapidly followed by coalescence, and relaxation towards an elliptical
profile. Total times scales to coalescence range from $\sim7\times10^{8}$
years to $\sim2\times10^{9}$ years, depending on the parameters of the
encounter. Gas and dust can be channeled into the nuclear regions of the
progenitors in one of two ways; either before coalescence, when tidal
forces during the first close approach form bars in one or both progenitors
(particularly if the progenitors have small bulge components) which are
very efficient at channeling material into the central regions of a galaxy,
or during coalescence, when shocks drive very large quantities of gas and
dust into the nuclear regions. More recent N-body simulations have in
general supported these conclusions, but added an intriguing possibility;
under certain conditions (particularly if the progenitors are gas
dominated), a merger between two disk galaxies can result in a disklike
remnant, rather than an elliptical \citep{naa03,spri05,rob05}.  It seemed
therefore that galaxy mergers could provide all the necessary physical
conditions for triggering a ULIRG, and that this implicated ULIRGs in the
formation of elliptical galaxies. Furthermore, these simulations could
explain a result that had been puzzling; that ULIRG activity could
apparently be triggered in mergers when the progenitors are still
physically separate.

\begin{figure}
\centering
\includegraphics[height=12cm]{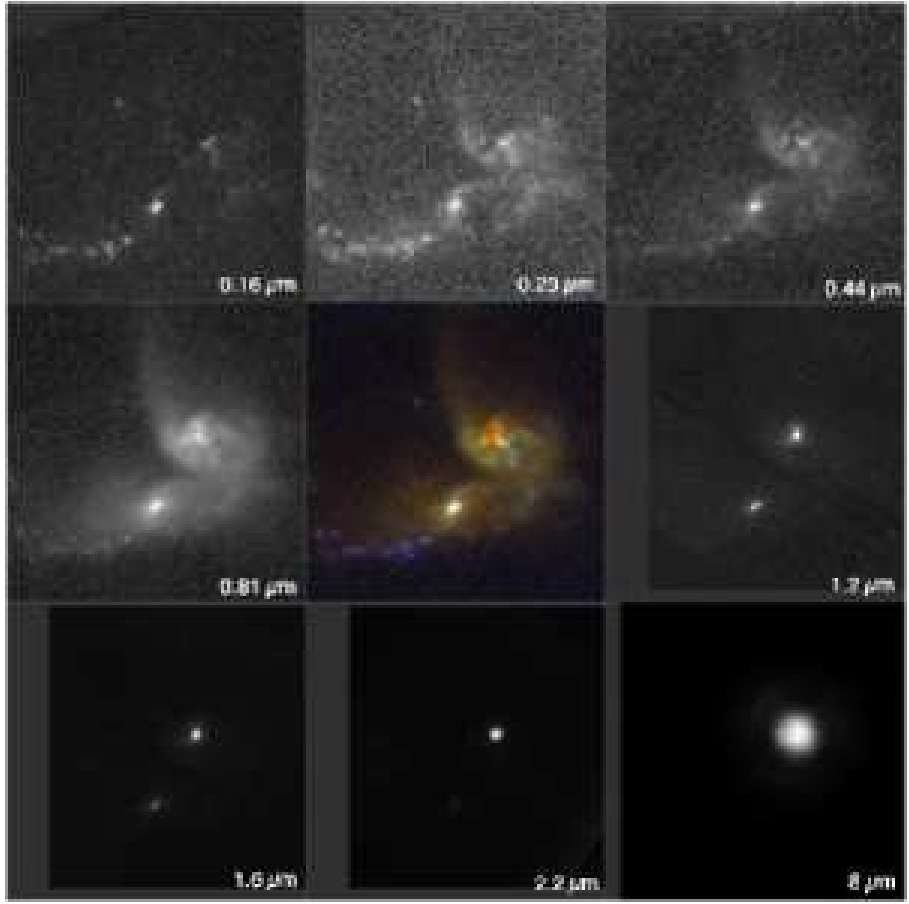}
\caption{IRAS 08572+3915, as seen at many wavelengths. The colour composite
at the center encodes wavelengths from the far-UV to the
near-IR as blue-to-red, and illustrates the complex composite nature of
ULIRG systems. The ultraviolet emission is
dominated by one of the two merger nuclei and by young super-star clusters
along the leading edge of one of the tidal tails.  Almost no ultraviolet or
blue emission is seen from the NW nucleus. At optical wavelengths,
increasingly dominated by old stars, we see the merger galaxy body, which
appears to be two spirals, accompanied by tidal tails. The NW nucleus has a
complex structure of dust lanes. As we proceed into the thermal infrared,
all emission sources other than a single compact emission source in the NW
nucleus fade away. The images from 0.16-2.2$\mu$m are from
STIS, WFPC2, and NICMOS \citep{gold,sur,sur00,sco}. The 8$\mu$m image is from Spitzer (Surace et al, in
prep), and has a beam size $\sim 5$ times greater than the HST data, but is
sensitive enough to exclude much emission from the SE nucleus. Keck
observations \citep{soi00} at similar wavelengths constrain the size of the
nucleus to $<0.3$ arcseconds, or 300 pc.  (J. Surace, priv. comm.).}
\label{ulirgimage}       
\end{figure}

Optical imaging from the Hubble Space Telescope (HST) offered
enhanced resolution and sensitivity over ground based facilities. An early
HST study of ULIRGs using WFPC2 \citep{sur} focused on a small sample with
`warm' infrared colours\footnote{i.e systems with $f_{25}/f_{60} > 0.2$,
where $f_{25}$ and $f_{60}$ are the 25$\mu$m and 60$\mu$m IRAS fluxes
respectively}, which biases towards systems likely to contain an obscured
AGN. All of the sample were found to be interacting, with complex
structures in their nuclear regions. Several systems showed a large number
of compact bright `knots' a few hundred pc in diameter, whose ages suggest
they result from the merger. A complementary ground-based survey of ULIRGs
with `cool' IR colours \citep{sur00} found a similar picture; all the
galaxies show signs of interactions, from early to late stage, and many
systems harbour `knots' similar to those seen in the warm sample. The
optical magnitudes are in most cases relatively modest (at least compared
to the enormous IR luminosities); most systems have $\sim L^{*}$
luminosities, with very few being substantially brighter. A later HST
survey of a larger, unbiased ULIRG sample \citep{far1} found similar
results; nearly all of the sample are interacting, with a wide range of
merger stages. A small number host optical QSOs whose host galaxies are
either interacting, or elliptical-like. Other authors have, on the basis of
HST data, suggested that some ULIRGs show evidence for mergers between more
than two galaxies, suggesting that ULIRGs may be the remnants of compact
galaxy groups \citep{bor00}. HST imaging of ULIRGs in the near-infrared
with NICMOS \citep{col01,bus2002} has produced very similar results to
those from optical imaging, with at least 90\% being interactions between
two or (possibly) more progenitors over a wide range of merger stages, and
that very few are much brighter than $L^{*}$. 

To date however, the largest optical/NIR
imaging survey of ULIRGs has been done from the ground; \citet{vei3}
present R and K images for 118 ULIRGs from the IRAS 1Jy spectroscopic
survey, and find that virtually all show signs of interactions, though very
few showed definite signs of interactions between more than two
progenitors. Most of the systems appear to be late-stage mergers,
especially for the more luminous systems and/or those with spectroscopic
signatures of an AGN, and typically have luminosities (in the K band) of
$L^{*}$ or greater. The most advanced mergers show evidence for emerging
elliptical profiles. 

Near diffraction-limited Keck mid-infrared imaging of ULIRGs has been
obtained by Soifer et al. (1999; 2000) who find extremely compact
structures, with spatial scales smaller than 0$^{\prime\prime}$.3 in six of the seven
ULIRGs observed. These compact sources emit between 30\% --- 100\% of the
mid-infrared energy from these galaxies. In Mrk 231, IRAS 05189-2524 and
IRAS 08572+3915 there is strong evidence that the source size increases
with increasing wavelength, suggesting heating by a central rather than
extended source, consistent with the optical classification as an AGN.
Spitzer mid-IR imaging programs of over 200 nearby ULIRGs (ongoing programs
of J. Surace (Fig. 2) and J. Mazzarella) will provide great sensitivity in
this important wavelength region, if limited spatial resolution.

\subsection{Optical \& near-IR Spectroscopy} 

It might be suspected that spectroscopy at wavelengths shortward of a few
microns would be useful mainly for redshift surveys, given the
dust-enshrouded nature of ULIRGs. This, however, is a misconception;
optical spectroscopic signatures of starburst and AGN activity are still
apparent even when such activity is moderately obscured (up to and
surpassing $A_{V}{\sim}10$ depending on the observations), and starbursts and
AGN can be detected in polarized light even when the direct line of sight
is completely hidden. Furthermore, spectroscopic diagnostics from the UV
through to the near-IR are, on the whole, more mature than those at longer
wavelengths \citep{bpt,vei87,ost,dop}, and line
strengths, ratios and profile shapes can provide powerful constraints on
the nature of the source of excitation. Recent advanced spectral synthesis
codes (e.g. Leitherer et al., 1999; Kewley et al., 2001) allow for
insightful diagnostics of starburst events, particularly in the UV where
direct emission from hot young star photospheres, rather than reprocessed
light from dust, is being sampled.

Early optical spectroscopic surveys of ULIRGs generally showed that they
were mostly starburst-like in the optical \citep{els85}, while samples with
`warmer' infrared colours appeared more biased towards Seyferts or LINERS
\citep{deg85,ost85}, or even optical QSOs \citep{bei86}. Conversely,
studies of ULIRGs with `cool' IR colours generally indicated the presence
of starbursts \citep{hek87,arm88}, sometimes accompanied
by spectral signatures of Wolf-Rayet stars, indicating that the starburst
was likely only a few Myr old. Also noted at this time was a tendency for
more IR-luminous objects to exhibit Seyfert spectra, with narrow lines in
direct light (Cutri et al., 1994) and broad lines in polarized light
(e.g. Hines et al., 1995).

To gain a complete picture of the spectroscopic properties of ULIRGs,
however, requires large-scale surveys, which were soon forthcoming as part
of IRAS followup. One of the largest to date is the
IRAS 1Jy survey \citep{kim98,vei99a}, which showed that the majority of
ULIRGs have optical spectra reminiscent of starbursts, but with a systematic
increase in the fraction of ULIRGs with Seyfert (1 or 2) spectra with
increasing IR luminosity.  Most of the ULIRGs with Seyfert spectra however
also show evidence for ongoing or recent star formation. Approximately
30\% of the Seyferts are Sy1's, with a systematic increase compared to
Sy2's with IR luminosity. Other notable findings included that optical
reddening generally decreases with increasing distance from the nuclear
regions and that the optically derived star formation rates are in most
cases many times lower than those derived from mid-/far-IR data. Followup
spectroscopy in the near-IR \citep{vei97,vei99} refined and extended this
picture, showing that, overall, around 25\% of ULIRGs show evidence for
an AGN and that this fraction increases with increasing IR luminosity
(reaching $\sim50\%$ at $L_{ir}>10^{12.3}L_{\odot}$). Those ULIRGs with
`warm' IR colours are more likely to show broad lines in the near-IR than
`cool' ULIRGs, and there is no observed correlation between extinction in
the NLR and the presence of broad lines in the near-IR, suggesting that the
NLR and BLR do not lie along the same line of sight. An intriguing further
result is that, of all the objects that show broad lines in the near-IR,
all are Sy2's in the optical, with no LINERs or HII's. Furthermore, most
($\sim70\%$) of the ULIRGs that show a Sy2 spectrum in the optical show
broad lines in the near-IR.

Very recent spectroscopy of ULIRGs from the UV through to the near-IR has
revealed some further important details. High spatial resolution UV and
optical spectroscopy using the Space Telescope Imaging Spectrograph onboard
HST of four `warm' ULIRGs \citep{far2005} has shown that the `knots' seen
in optical imaging in many cases harbour very luminous starbursts and AGN,
implying that these optically bright knots may also be the sites of the
heavily obscured power sources behind the IR emission. The spectral
properties of some of these knots also suggest further links between ULIRGs
and (low ionization) broad absorption line (BAL) QSOs, and the forming
cores of elliptical galaxies. The starbursts in these knots were all
observed to be young, with ages of 4Myr-20Myr, supersolar metallicities,
and an IMF (Salpeter) slope of less than about 3.3. Near-IR spectroscopy
with the Very Large Telescope (VLT) in Chile, and with Keck
\citep{gen01,tac2002} has shown that the host galaxy kinematics of ULIRGs
resemble those of ellipticals with luminosities of $\sim L^{*}$ (but with a
large scatter), and that the host galaxy properties of those ULIRGs that
contain an AGN and those without are very similar.

Finally, there is strengthening evidence that nuclear and galactic scale
outflows may be common in ULIRGs \citep{hek87, wil99}, and high resolution
spectroscopy has discovered galactic-scale outflows with high ejection
efficiencies in many ULIRGs that are dominated by star formation
\citep{rup02}.  \citet{lip03} noted that relatively low velocity outflows
are present in starburst dominated ULIRGs, but that higher velocity
outflows are present in systems that show evidence for both a starburst and
an AGN, though even in composite systems,
the starburst is still probably the dominant mechanism behind
galactic-scale outflows \citep{rup05}.

\subsection{Mid-infrared spectroscopy}

An area where the Infrared Space Observatory, ISO \citep{kes}, excelled in
the study of ULIRGs was in spectroscopy of nearby systems using SWS
\citep{degraauw}, LWS \citep{clegg}, ISOCAM \citep{ces} and ISOPHOT-S
\citep{lemk}, because for the first time high-sensitivity and
high-resolution spectroscopy could be obtained in a wavelength region with
minimal effects of dust extinction, at least compared to the optical.
Moreover the mid-infrared region holds several diagnostic lines which are
very useful for characterization of the major power source for ULIRGs.  The
mid-infrared continuum shape can be a powerful diagnostic itself, betraying
the existence of warm dust in the close vicinity of an AGN.  We refer the
reader to several previous reviews on ISO's legacy on star forming
galaxies, AGN and ULIRGs \citep{genc,elb2,ver,olivpozz} for more details.

\citet{gen} demonstrated the power of SWS spectroscopy to separate AGN from
starbursts by comparing the strength of the 7.7$\mu$m PAH equivalent width
to high/low excitation line ratios, such as
[OIV]25.9$\mu$m/[NeII]12.8$\mu$m; PAH molecules will be destroyed by
high-intensity AGN radiation fields so low PAH/continuum ratios should
correlate with strong [OIV]/[NeII] ratios.  They concluded that at leat
half of their sample of 15 local ULIRGs have simultaneous starburst and AGN
activity, but 70-80\% of the sample is predominantly powered by star
formation (a result mirrored by mid/far-IR SED fitting, \citet{far2}).  The
method requires high sensitivity to determine the fine structure line
ratios, and the PAH emission can also be contaminated by strong silicate
absorption at 9.7$\mu$m. Moreover dense nuclear environments can hide AGN
activity even at mid-IR wavelengths, as discovered by \citet{cla} in
Seyfert 2 nuclei in which the the mid-infrared continuum from the AGN is
sufficiently absorbed to allow extranuclear PAH emission to dominate the
spectrum.  \citet{lau0} developed a diagnostic diagram based on the
6.2$\mu$m PAH feature relative to continuum strength versus mid-IR
continuum colour, which could be used for systems lacking fine-structure
line spectroscopy, and \citet{pee} extended this approach to include
far-infrared continuum colours.
\citet{lut0,rig} and \citet{tra} applied these techniques to large ULIRG
samples studied with ISOPHOT-S, concluding, in agreement with \citet{gen},
that starbursts predominantly power these systems, though the presence of
heavily obscured AGN cannot be ruled out.  

Soifer et al. (2002) obtained low spectral resolution, but high angular
resolution, Keck mid-IR spectra of five LIRGs---ULIRGs demonstrating that
PAH emission, when present, generally is circumnuclear in origin, extended
over scales of 100-500pc.  The silicate optical depths in these sources can
be as high as 15, suggesting that even mid-IR spectroscopy may not be
probing the true nuclei in the most compact sources.

Mid-IR spectral classifications, however,
generally agree with optical line classifications of AGN vs star formation
power, although some ULIRGs with LINER-like optical spectra were
interpreted as starburst dominated in the mid-IR, attributed to starburst
wind-driven ionising shocks \citep{lut99,sug} instead of to low-level AGN.
However, the LINER situation is complex as demonstrated by subsequent
comparison of ISO-SWS fine structure line spectroscopy with Chandra X-ray
imaging of a sample of LINERs. \citep{say} has shown that LINERs are
intermediate between starbursts and AGN in mid-infrared line excitation,
and that most LINERs contain a compact hard X-ray source characteristic of
an AGN.  They also found some anti-correlations between mid-IR fine
structure line diagnostics and hard X-ray AGN diagnostics: their highest
excitation mid-IR spectrum source NGC 404 shows only weak soft X-ray
emission while the low mid-IR excitation LINER NGC 6240 shows an extremely
luminous binary X-ray AGN (Komossa et al., 2003). The most likely explanation
for objects of this kind are extremely high optical depths even at
mid-infrared and X-ray wavelengths, and potentially different
lines-of-sight to the AGN core at different wavelengths \citep{ris}, and/or
unusual ratios of gas-to-dust optical depths.

\citet{stu} have developed diagnostic diagrams using [OIV]25.9$\mu$m,
[SiII]34$\mu$m, [NeVI]7.65$\mu$m and [NeII]12.8$\mu$m and other fine
structure lines, coupled with Brackett $\beta$ at 2.63$\mu$m, to separate
high from low excitation systems and estimate the fractional contribution
of star formation and AGN to the integrated light using mixture lines, in
analogy to the well-known optical line diagnostic diagrams \citep{vei87}.
Most ULIRGs were not detectable by ISO in these lines, though
Arp 220 was shown to exhibit low excitation in these diagrams.

\citet{spi} proposed an analogous far-infrared approach using
[CII]158$\mu$m, [OI]63$\mu$m and [OII]88$\mu$m, though again very few
ULIRGs were detectable in these lines by ISO.  Arp 220 has been observed
extensively by ISO, as the nearest and brightest ULIRG. Arp 220's FIR
spectrum is dominated by molecular absorption from species common to
Galactic photo-dissociation regions (PDR): OH, H$_2$O, CH, NH, NH$_3$
\citep{fis} and is very weak in the fine structure emission lines, with
[OI] 63$\mu$m in absorption.  The [CII]158$\mu$m fine structure transition
decreases with increasing IR luminosity (and more strongly with FIR colour
--- $S_{60\mu m}/S_{100\mu m}$) in infrared galaxies \citep{mal}, and in
ULIRGs is only 10\% of that measured in less luminous starbursts
\citep{fis,luh}. Since [CII] is the primary coolant in the ISM of normal
galaxies it was expected to be strong in high-star formation systems, and
its apparent lack is interpreted to indicate a stronger radiation field in
luminous starbursts producing charged grains which result in a lower
heating rate in PDRs \citep{mal01}.  \citet{gon04} have modeled the FIR
spectrum of Arp 220 over the spectral range 40-200$\mu$m.  Their model
requires three components: the compact nuclei, modeled as a unit, with
$T=10^6K$ which are optically-thick throughout the infrared, an extended
region which is dominated by PDR emission at $T \sim 40-90K$, and a halo
which produces absorption from the low-lying levels of OH and H$_2$O plus
the CH.  Given that the two nuclei show distinct mid-IR spectra
\citep{soi99} it is unclear what effect modeling the nuclei as a single
source may have.  In this model the [CII] emission is produced in the
extended (PDR) region with little [CII] emission from the nucleus, which
has distinctly non-PDR conditions.

For the majority of ULIRGs for which spectroscopy is not available
broadband colours have been used to characterize the broad spectral type and
major infrared energy source, following the early IRAS schemes
\citep{kla,spi,far2}. ULIRGs are found to fall into two general classes by
these methods: "cool" systems dominated by star formation, which show a
relatively flat mid-infrared spectral shape and a steep rise towards longer
wavelengths, and "warm" systems with a red power-law-like SED through the
mid-infrared, usually assumed to be AGN-dominated, although compact HII
regions can also show a warm mid-IR SED (Dopita et al., 2005).  Warm ULIRGs
tend to have Seyfert 1 like optical spectral features while cool ones have
starburst or Seyfert 2 optical spectra.  The two clases were found to be
indistinguishable in the far-infrared \citep{kla}, suggesting that at these
wavelengths the emission may not be dominated by an AGN in either class,
however \citet{pee} find ULIRGs to have more prominent far-infrared
emission than AGN or lower luminosity starbursts such as M82.

Several lines of evidence indicate that AGN become more bolometrically
significant with increasing IR luminosity in ULIRGs. The incidence of both
AGN optical line indicators and mid-infrared continuum and line AGN
diagnostics were found to increase with increasing infrared luminosity
\citep{shi,gen,lut0,rig,tra}. However, these ISO HLIRG samples are biased
towards previously known AGN so Spitzer confirmation of these result is
important, using unbiased samples of infrared-selected HLIRGs.

\begin{figure}
\includegraphics[height=10cm]{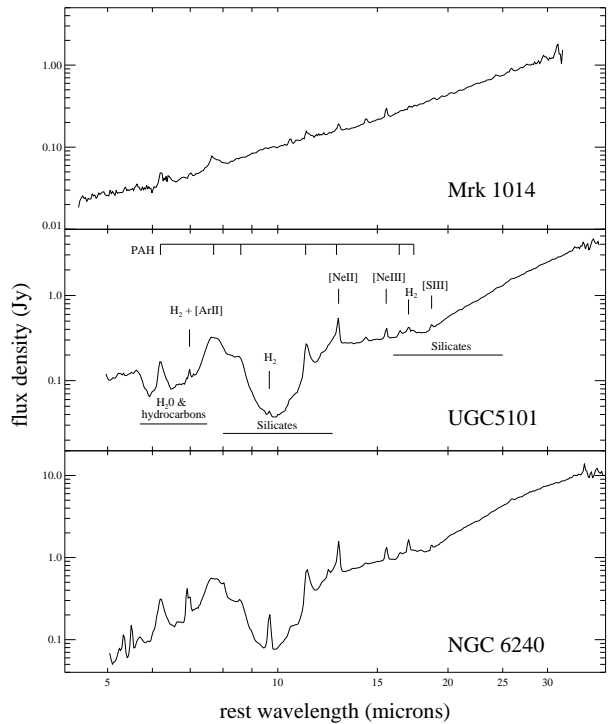}
\caption{IRS low-resolution spectra of Mrk 1014, UGC 5101 and NGC 6240.
The positions of prominent emission features and absorption bands (the 
latter indicated by horizontal bars) are marked on the UGC 5101 spectrum.  
Figure courtesy of L. Armus and J. Houck.}
\label{irsplots} 
\end{figure}

Observations with the ISO satellite greatly expanded our understanding of
the mid-infrared spectra of ULIRGs, however, many ULIRGs were beyond the
reach of many of the diagnostic methods until the advent of Spitzer.
Several extensive Spitzer programs are underway to adequately sample the
local ULIRG population, the most comprehensive being a study of the
mid-infrared spectra of a large number ($ > 100$) of ULIRGs having $0.02 <
z < 0.93$ with the Infrared Spectrograph (IRS), as part of the IRS
guaranteed time program.  These sources are chosen primarily from the IRAS
1-Jy (Kim \& Sanders 1998), 2-Jy (Strauss et al., 1992), and the FIRST/IRAS
radio-far-IR sample of Stanford et al. (2000).  In Figure \ref{irsplots} we
show IRS spectra for three nearby ULIRGs (Mrk 1014, UGC 5101, and NGC 6240)
whose spectra serve to highlight the range in properties seen throughout
much of the sample, and the power of the IRS in wavelength coverage and
sensitivity for studies of the nuclei and interstellar media in dusty
galaxies (Armus et al., 2004; 2005).

Mrk 1014 ($z=0.1631$) is a radio-quiet, infrared luminous QSO with broad
optical emission lines and twin tidal tails indicative of a recent
interaction (MacKenty \& Stockton 1984).  UGC 5101 ($z=0.039$) has a
single, very red nucleus within a disturbed morphology suggestive of a
recent interaction.  Optically, UGC 5101 is classified as a LINER (Veilleux
et al., 1995).  It has a high brightness temperature ($T>10^{7}$K) radio
nucleus at 1.6 GHz which is resolved with the VLBA (Lonsdale et al., 1995).
ISO SWS and PHT-S spectroscopy (Genzel el al., 1998) indicate a powerful,
circumnuclear starburst.  Based upon its IRAS colours, UGC 5101 is
classified as a cold, starburst-dominated, far-infrared source.  However,
XMM data indicate an obscured, but luminous, hard X-ray source with
$L_{x}$(2-10 keV)$\sim 5\times 10^{42}$ erg s$^{-1}$ and $L_{x}$(2-10
keV)/$L_{IR} \sim 0.002$ suggestive of a buried AGN (Imanishi et al.,
2003).  NGC 6240 ($z=0.0245$), is a double-nucleus, merging galaxy (Fosbury
\& Wall 1979), with an $8-1000\mu$m luminosity of $\sim 7\times 10^{11}
L_\odot$.  The optical nuclear spectrum of NGC 6240 is classified as a
LINER (Armus, Heckman \& Miley 1987), and the extended optical nebula
reveals the presence of a starburst-driven superwind (Heckman, Armus \&
Miley 1987).  X-ray observations with ASCA (Turner et al., 1998), Beppo-Sax
(Vignati et al., 1999), Chandra (Komossa et al., 2003; Ptak et al., 2003),
XMM-Newton (Netzer et al., 2005) provide clear evidence for the presence of
one (or two) AGN behind significant columns of absorbing material ($N_{H} =
1-2\times 10^{24}$ cm$^{-2}$).  Relatively strong [OIV] $25.89\mu$m line
emission in the ISO SWS spectrum of NGC 6240 led Lutz et al. (2003) to
suggest that up to $50\%$ of the infrared energy emitted by NGC 6240 could
be powered by a buried AGN.

The spectra of Mrk 1014, UGC 5101, and NGC 6240 are strikingly different.
Mrk 1014 has a steeply rising mid-infrared spectrum with weak emission
features and little or no silicate absorption.  The spectra of UGC 5101 and
NGC 6240, on the other hand, are dominated by strong silicate absorption at
$9.7\mu$m and $18\mu$m, and PAH emission at 6.2, 7.7, 11.3, and $12.7\mu$m
(Armus et al. 2004, 2005).  The extinctions toward the nuclei in UGC 5101
and NGC 6240, as estimated from the depths of the silicate absorption, are
at least A$_{V} = 15-35$ and A$_{V} = 60$ mag, respectively.  UGC 5101 also
shows strong absorption between $5-7.5\mu$m from water ice and
hydrocarbons.  As suggested by Spoon et al. (2002), the water ice features
may indicate the presence of shielded molecular clouds along the line of
sight to the nucleus.  NGC 6240 has little or no water ice absorption, but
very strong emission lines from warm ($T\sim300-400$K) H$_{2}$.  The mass
of this warm gas is estimated to be approximately $1.4\times 10^{8}
M_\odot$ -- about $1\%$ of the cold molecular gas mass derived from
single-dish millimeter CO line measurements (Solomon et al. 1997), but up
to $3-7\%$ of the cold molecular gas within the central 1 kpc measured by
Tacconi et al.  (1999).

The IRS high-resolution (R=650) spectra (not shown) provide important
diagnostic measures of the dominant ionizing sources in the ULIRGs
because it is possible to accurately measure unresolved atomic, fine-structure
lines of Ne, O, Si, and S, covering a large range in ionization
potential.  The [NeV] 14.3 / [NeII] 12.8 and [OIV] 25.9 / [NeII] 12.8
line flux ratios in Mrk 1014 (0.9 and 1.7, respectively) suggests that
nearly all the ionizing flux comes from the central AGN, although the
obvious presence of PAH emission suggests some extranuclear star
formation.  In UGC 5101 and NGC 6240 weak [NeV]
emission has been detected, with $14.3\mu$m line fluxes of $5-6 \times 10^{-21}$W
cm$^{-2}$, indicating buried AGN.  While the [NeV] / [NeII] and the
[OIV] / [NeII] line flux ratios imply an AGN contribution of $< 10$\%
to the total luminosity in both sources (Armus et al. 2004, 2005), the
large optical depth to the nuclei, as evidenced by the deep silicate
absorption and X-ray columns, leaves open the possibility that the
true contribution of the AGN to the bolometric power output in UGC
5101 and NGC 6240 may be much larger than revealed by the mid-IR
emission lines.  In NGC 6240, the extinction-corrected hard X-ray data
are consistent with the buried AGN producing $50-100\%$ of the
luminosity.  An inclined, dusty torus, patchy extinction, and/or a low
covering factor for the [NeV]-emitting clouds, could reconcile these
apparently discrepant estimates -- an explanation often evoked to
explain observations of other type-2 AGN.

Spoon et al. (2004) observed the more distant ULIRG IRAS
F00183-7111 at z=0.327, detecting strong absorption from CO$_2$ \& CO gas,
water ice, hydrocarbons and silicates, indicating high obscuration, a
complex line of sight, and the presence of very warm, dense gas (720K).
Direct signs of an obscured AGN are not found but Spoon et al. conclude
that an obscured AGN probably accounts for most of the luminosity based on
similarities to other systems with highly obscured AGN.
 
\subsection{Radio Continuum Studies: AGN {\it vs} Luminous Radio Supernovae}

The existence of a tight correlation between integrated far-infrared
flux-density and radio continuum emission --- the "Radio-FIR Relation"
\citep{hel,yun} --- over four orders of magnitude in IR luminosity, allows
the use of high-resolution radio interferometric techniques to study the
compact nuclei of luminous IR galaxies in one of the few spectral regions
with relatively low optical depth.  The Radio-FIR relation is not well
understood, but is believed to be produced by nonthermal radiation from
relativistic electrons in, or leaking out of, starburst-related supernova
remnants.  The seminal study of a complete sample of luminous IR galaxies
by Condon et al. (1991) with the VLA at 1.49 and 8.44GHz demonstrated the
extreme compactness of luminous IR nuclei and concluded that most of their
40 galaxies -- 6 of which are ULIRGs -- are consistent with a
starburst-related origin for the radio emission.

\citet{lon93} used a global VLBI array to detect 18cm high brightness
temperature, $T_b \ge 10^6K$, emission cores from 17 of 31 luminous (log
$L_{FIR} \gtrsim 11.25 (L_\odot)$) infrared galaxies, consistent with such
cores existing in all such LIRGs at $\sim$10\% of the total 18cm flux
density.  This result indicates either obscured radio-quiet AGN in most
LIRGs and ULIRGs, possibly energetically dominant \citep{lon95}, or clumps
of starburst-related luminous radio supernovae (LRSN) and remnants
\citep{smi1}.  Similar studies in the South \citep{nor,kew} detected
compact nuclear radio cores with lower frequency, but their results are
consistent, allowing for differences in sample selection and sensitivity.

Subsequent VLBI imaging dramatically revealed a cluster of luminous RSN in
the nearest ULIRG, Arp 220 \citep{smi1} --- clear evidence for a high star
formation rate, estimated at that time as ${\sim} 100 M_{\odot}$, yr$^{-1}$ by
\citet{smi0}, which would be enough to power the infrared luminosity by
star formation without resorting to an obscured AGN.  \citet{momj,neff,bon}
have reported possible "supernova factories" in IRAS 17208-0014, Mrk 299
and Mrk 273. respectively. The original dozen LRSN in Arp 220 have been
monitored with VLBI \citep{rov05} and more sensitive VLBI imaging studies
\citep{lon06} have revealed a number of new and fainter LRSN.  These
Luminous RSN were originally modeled after RSN1986J in NGC 891 \citep{wei},
which exhibited a maximum radio power, log $P_{1.5GHz}^{1986J} = 21.15$.
The LRSN in Arp 220, however, exhibit much slower decay in their radio
light curves \citep{rov05} than RSN 1986J; longer RSN lifetimes reduce the
inferred star formation rate, therefore it is not clear from the VLBI
imaging whether a starburst is the chief power source for Arp 220.

On the other hand \citep{lon2003} demonstrated that the radio emission in
UGC 5101, Mrk 231 and NGC 7469 is AGN dominated.  Mrk 231, in some senses
the classical "infrared quasar", has been studied with the VLBI by
\citet{ulv}.  Their images (see also Lonsdale et al. 2003a) show a triple
structure, with a core and two lobes which classify it as a Compact
Symmetric Object (CSO).  It has been suggested that CSOs are young, $\tau
<< 10^6$yr, with the hot spots representing the working surface of a
relativistic jet upon the ambient medium \citep{read96}. If the southern
(primary) lobe/hot-spot in Mrk 231 is confined by ram pressure,
\citep{lon2003} estimate a lobe advance speed, $v_a \sim 10^{-4}c$ and an
age for the jet/compact source, $\tau < 10^6$ yr.  Despite the clear
evidence for AGN domination of the radio structure in these systems, the
radio power is small compared to the infrared/bolometric emission, which
may still therefore be starburst dominated.  In Mrk 231, for example,
several studies \citep{car,lon2003,far2} suggest that more than half of the
total luminosity comes from a circumnuclear starburst in a molecular ring
rather than from the AGN.

\subsection{Molecular gas: CO \& HCN Observations}

The interpretation of Luminous Infrared Galaxies as starburst systems was
strengthened soon after their identification with IRAS by studies of
neutral hydrogen \citep{ms88} molecular gas, principally CO
\citep{san86,san91} but more recently HCN \citep{gs04}
and OH maser emission \citep{baa,baa1}. These early
studies demonstrated that ULIRGs as a class exhibit compact nuclear
reservoirs of high-density gas, and, with mass estimates of order
$10^9-10^{10}M_\odot$ in HI and in H$_2$, are consistent with the
interpretation that star formation accounts for a substantial fraction of
the FIR luminosity in Luminous IR Galaxies.
                                     
Millimeter-wave interferometer measurements of CO emission (e.g. Scoville
et al. 1991; Solomon et al. 1997; Downes \& Solomon 1998; Bryant \& Scoville
1999) have demonstrated that nearly half of the CO mass in Luminous IR
Galaxies is contained within the central regions, $r < 0.5-1$ kpc, with as
much as $10^{10}\, M_{\odot}$ of molecular gas distributed in nuclear disks
of radius a few hundred pc, but thicknesses of one-tenth the radius and
densities over 10$^4$ cm$^{-3}$ \citep{bs99}. The mean
molecular surface densities in these structures may exceed $10^4 M_{\odot}$
pc$^{-2}$ with the molecular gas providing a large fraction of the dynamical
mass.  The picture of molecular gas in ULIRGs is very different from that
in the Galaxy, with much higher surface densities and inferred optical
depths ($A_v \gtrsim 10^2 - 10^3$).  Early CO studies assumed a Galactic
conversion ratio between CO Luminosity and H$_2$ mass, $M_{H_2}/L^\prime_CO =
4.6\, M_\odot$ (K km s$^{-1}$ pc$^{-2})$, however, recent analyses suggest
that this may overestimate the molecular mass, either owing to
high-brightness temperatures in the CO emission ({\it e.g.} Mrk 231; Bryant
\& Scoville 1999) or because the line width/velocity dispersion reflects
the total dynamical mass in the central nuclear region, rather than that
within virialized molecular clouds \citep{sol97}.

Again, Arp 220 provides a convenient laboratory to study molecular gas, if
not always in a typical ULIRG environment.  \citet{sak99} have
studied the Arp 220 nuclei in the CO(2-1) transition and accompanying 1mm
continuum.  Steep velocity gradients are found in the CO associated with
each of the two nuclei, which are misaligned with each other and with the
outer CO disk.  Sakamoto et al. interpret these as molecular disks
associated with the merging nuclei, counter-rotating with respect to each
other and with respect to the outer disk.  The dynamical masses inferred
from CO kinematics are of the order of $2 \times 10^9 M_\odot$. The central
molecular gas is inferred to have high filling factor, more like a uniform
disk than individual clouds, which will have an inward accretion rate of
approximately $100 M_{\odot}$ yr$^{-1}$ \citep{sco97} which
is similar to the star formation rate of Arp 220 inferred from the FIR
Luminosity \citep{smi0,far2}.
                                  
Early studies placed considerable emphasis on the ratio of FIR Luminosity
to CO Luminosity --- often called the Star Formation Efficiency --
demonstrating that the ratio of FIR/CO increases at higher $L_{FIR}$.  This
is interpreted as either an increase in Star-Formation Rate per unit
molecular gas mass, {\it i.e.} more efficient star formation, or as
possible evidence for an alternative, AGN, contribution to the FIR
luminosity (Sanders \& Mirabel 1996).  More recently, studies of HCN,
which, owing to the higher dipole-moment of the HCN molecule is a tracer of
warmer ($T_{kin} \sim 60-90$K), higher-density ($n_{H_2} \sim 10^5 -
10^7$ cm$^{-3}$) gas, shows a much higher ratio of HCN/CO luminosity in
Luminous IR Galaxies than in quiescent spirals like the Milky Way
\citep{sol92}.  Furthermore the FIR/HCN luminosity (HCN Star Formation
Efficiency, as above) shows an approximately linear relationship in
Luminous IR Galaxies with $L_{fir}$ from $10^{10}$ -- $10^{13}\,L_\odot$
indicating the presence of an abundant warm, high-density molecular
environment within ULIRGs \citep{gs04}.  ULIRGs thus have a much greater
dense, warm medium like that of the star-forming cores of Galactic
molecular clouds --- perhaps the central regions of the molecular disks,
although little spatial information is available about the HCN distribution
in ULIRGs.  Since there is little doubt that the lower luminosity systems
amongst this sample are dominated by star formation, the continuity of this
relation up to luminosities characteristic of HLIRGs is one of the most
compelling pieces of evidence, albeit statistical, that most ULIRGs are
dominated by starburst power rather than AGN power.

\subsection{Maser Emission}

Luminous IR Galaxies have been known to be strong emitters of OH maser
emission since the discovery of OH 1667MHz emission in Arp 220 by
\citet{baa}, who dubbed it a "megamaser", having an OH luminosity roughly
$10^6$ times that of masers in the Galaxy. H$_2$O maser emission, which
has been used to map torus/disklike structures around compact supermassive
AGN nuclei such as NGC 4258 \citep{miy96} is not generally detected
in classical Luminous IR Galaxies (see Lo 2005 for a review).
   
OH megamasers are found preferentially in the most luminous IR galaxies
\citep{baa1}, and a rough dependence of the OH luminosity on the square of
the FIR luminosity was originally found, suggesting that the FIR emission
provided the pumping mechanism to provide the necessary population
inversion ({\it e.g.} Henkel \& Wilson 1990) within a foreground molecular
screen amplifying the diffuse radio continuum, which is in turn correlated
with the FIR emission.  The OH gas may trace very high density regions
(n$_{H_2} = 10^{5-7}$cm$^{-3}$), although an apparently separate component of the OH
megamasers is associated with a much lower-density high-velocity outflow
\citep{baa2}.
 
The initial interpretation of the maser phenomenon of a foreground, low-gain
molecular screen amplifying the diffuse radio continuum in Luminous IR
Galaxies, however, has been questioned by the VLBI observations of Arp 220
\citep{dia,lon94,lon98,rov} demonstrating that over two-thirds of the
masing gas in Arp 220 is compact, produced in structures a few pc$^3$ in
volume with amplification ratios of order 10$^3$ or higher.  These compact
masers have complex spatial and velocity structure arising in clouds
generally within the Arp 220 nuclei, but not coincident with nuclear radio
continuum (or any other detectable radio continuum).  The southern
component within the western nucleus (the one with the majority of LRSN)
shows a velocity gradient in excess of 18,000 km\,s$^{-1}$ \citep{rov},
which, interpreted as rotation, implies a mass of order $2 \times 10^7
M_\odot$.
  
Similar compact maser emission has been detected with Global VLBI in
IIIZw35 and IRAS17208-0014 \citep{dia99}, and in 12032+1707 \citep{pil}. On
the other hand Mrk 231 \citep{lon2003,klo}, Mrk 273 \citep{klo2} and
14070+0525 \citep{pil} show more extended emission in a circumnuclear disk
with conditions more like the classical maser model. IIIZw35 also shows
diffuse ring of maser emission with radius $\sim 20$pc and central mass
$\sim 7\times 10^6 M_\odot$ \citep{pil01}. In this case the compact
emission occurs at the tangent points of the ring, and the maser structure
has been modeled by multiple high density clouds within the diffuse ring.
Such a structure does not, however, appear to be able to explain the
complex, compact maser emission in Arp 220.

In the case of Mrk 231, the OH emission appears to be the central portion
of a circumnuclear HI disk detected in absorption by \citet{car}, though it
is misaligned with the central CO distribution \citep{klo2}.
\citet{lon2003} speculate that the ignition of Mrk 231's Seyfert nucleus
may have disrupted compact maser emission in the nucleus itself.

An extensive survey for OH maser emission in over 300 IRAS galaxies from
the PSCz Survey \citep{sau} with $z > 0.1$ has been carried out with the
upgraded Arecibo Telescope by Darling \& Giovanelli (2002 and references therein).  This
survey has expanded the sample of known megamasers in luminous infrared
galaxies to over 100 with detections in nearly 20\% of the galaxies
surveyed.  Their analysis suggests few correlations between OH and FIR
properties.  A re-analysis of the $L_{OH} - L_{FIR}$ relationship in their
data suggests a much flatter relation $L_{OH} \propto L_{FIR}^{1.2}$
consistent with other recent analyses \citep{kan0} which finds a slope of
approximately 1.4, suggesting a mixture of extended, unsaturated emission
plus compact, saturated clouds.  One source in the \citet{dar} sample, IRAS
21272+2514, shows apparent variability which the authors interpret as due
to interstellar scintillation.  This interpretation requires $\sim$ 30-60\%
of the OH maser emission to originate in saturated clouds of dimension
smaller than $\sim$ 2pc.

A satisfactory model for the complexities of OH masers in ULIRGs has yet to
be proposed.  Almost certainly the diffuse emission follows the classical
model with largely unsaturated OH maser clouds, pumped by the strong FIR
radiation field, amplifying the nuclear 1.6GHz radio continuum.  The
compact, pc-scale structures in Arp 220 and other galaxies, coupled with
the lack of detectable associated 1.6GHz continuum emission, requires a
substantial saturated maser component.  These structures subtend
solid-angles too small to intercept sufficient FIR photons to excite the OH
molecules, suggesting that collisional excitation must be important
\citep{lon98}.  A tentative correlation between OH line-width and
X-ray luminosity \citep{kan} may indicate that X-ray heating of the
molecular gas plays a role in collisional excitation.The possible roles of
shocks and/or AGN activity remain to be explored.

\subsection{X-Ray Emission}

The importance of X-rays has been recognized not only because of the
diagnostic ability of the X-ray to discriminate between AGN and Starburst
emission \citep{rie88}, but also because models of the X-ray background
(XRB) require substantial populations of highly-obscured AGN at redshifts,
$z \sim 0.5-1.5$, to reproduce the observed XRB spectrum (Worsley et
al. 2005).  Rieke's early analysis from HEAO A-1 demonstrated that ULIRGs
are underluminous in the 2-10keV band compared to classical Seyfert
galaxies or QSOs.  It has only been recently with the Chandra and
XMM-Newton X-ray observatories that the implications of this early discovery
could be investigated in significant ULIRG samples with sufficient
resolution and signal-to-noise to provide reliable diagnostics of the
nature of the nuclear X-ray sources.
 
Further evidence of the weakness of X-ray emission in ULIRGs comes from a
ROSAT survey \citep{bol} of 323 ULIRGs which detected fewer than
10\%. Probably the most significant starburst-related discovery of the
ROSAT soft X-ray satellite is the detection of extended thermal outflows,
first in nearer, lower-luminosity starbursts, and then in Arp 220, dubbed
"superwinds" \citep{hek96}.  Similar thermal components appear to be common
in ULIRGs and they are described as "ubiquitous" in starbursts with Star
Formation Rates $> 10^{-1} M_{\odot}$ yr$^{-1}$ kpc$^{-2}$ in Heckman's
(2001) review; see also the recent review by \citet{vei4}.  The superwinds
are believed to be driven by supernova supplied kinetic energy with outflow
rates comparable to the star-formation rates in these galaxies.  Heckman
speculates that these superwinds may be the principal "polluters" of metals
and dust into the IGM.  Further Chandra studies \citep{mcd} of the Arp 220
superwind reveal extended, faint, edge-brightened, soft X-ray lobes outside
the optical galaxy out to a distance of 10-15 kpc.  Bright plumes inside
the optical isophotes coincide with the optical line emission \citep{col04}
and extend 11 kpc from end to end across the nucleus. The data for the
plumes cannot be fitted by a single-temperature plasma and display a range
of temperatures from 0.2 to 1 keV. There is a close morphological
correspondence between the H$\alpha$ and soft X-ray emission on all spatial
scales.
 
There have been three recent ULIRG X-Ray surveys with XMM-Newton
\citep{fra03} which observed 10 ULIRGs, and with Chandra \citep{pta03,teng}
which observed 8 and 14 respectively.  All of the ULIRGs surveyed were
detected, with X-ray luminosities typically $L_{2-10keV} < 10^{42} -
10^{43}$ erg s$^{-1}$.  These luminosities represent $< 1$\% of the
infrared luminosities in these systems, confirming that ULIRGs are much
less luminous in the X-ray than classical AGN.  Furthermore, the soft X-ray
emission from all systems is dominated by extended, thermal emission with
$kT \sim 0.7$keV, and is uncorrelated with IR luminosity.  In at least two
XMM-Newton systems the emission is extended on scales of 10s of kpc,
suggesting a superwind origin.  In at least 5 galaxies a hard X-ray
(2-10keV) component and/or the presence of 6.4keV Fe K-line emission
suggests the presence of an AGN which is not energetically dominant
\citep{fra03}.  Similarly, in the nearer Chandra sample of Ptak et
al. (2003) all galaxies exhibit hard components which are interpreted as
AGN sources.  5 galaxies in the Chandra samples which are classed as AGN
(including IRAS 05189-2524, Mrk 231, Mrk 273) show order of magnitude
greater X-ray luminosities than the Starburst ULIRGs (UGC 5101, IRAS
17208-0014, IRAS 20551-4250, IRAS 23128-5919) with NGC 6240 being an
intermediate case\footnote{Mrk 231, IRAS 17208-0014 and IRAS 23128-5919 are
included in both the \citet{pta03} and \citet{fra03} samples}.  As already
noted UGC 5101 and Mrk231 are believed to harbor obscured AGN based on
their Spitzer spectra.  Mrk 273, NGC 6240 and UGC 5101 exhibit Fe K-line
emission, but sufficiently weak that the authors argue against X-rays
reflected from a Compton-thick ($N_H > 10^{24}$ cm$^{-2}$) absorber as the
origin of the observed X-ray emission.  Teng et al. (2005) use hardness
ratios to estimate the X-Ray spectral properties of their fainter
galaxies. The photon indices for the combined Chandra samples peak in the
range $1 < \Gamma < 1.5$, with a tendency for Seyfert ULIRGs to have
steeper spectra, $\Gamma > 2$.  Although the X-ray properties of this
fainter sample are consistent with a Starburst origin, the presence of
Compton-thick AGN cannot be ruled out and may be expected in many cases
given the results  discussed above.

Although the situation is complicated, certainly some ULIRGs must harbor
luminous X-ray AGN, though perhaps behind large absorbing columns and thus
only visible in hard X-rays.  In Mrk 231 BeppoSAX revealed a highly
absorbed ($N_H \sim 2\times 10^{24}$ cm$^{-2}$) power-law component
\citep{bra} and analysis of XMM-Newton data indicates that below 10 keV
only scattered or reflected X-rays escape.  In a 40ks observation of
of Mrk 231, \citet{gal} find the majority of the X-ray luminosity is
emitted from an unresolved nuclear point source with a very hard spectrum,
the majority of the flux emitted above 2 keV. The source is also variable
on a timescale of a few hours.  They argue against a Compton-thick
reflection model of Mrk 231 \citep{malr}, proposing a Compton-thick
absorber which allows scattered light from multiple lines of sight to be
detected. Highly absorbed AGN power-law hard X-ray sources are also
reported in Mrk 273 \citep{xia,bal}, NGC 6240 \citep{iwa98,vig,net}, and
Mrk 1014 \citep{bol02} although the starbursts in these galaxies dominate
in softer X-rays, and possibly also bolometrically.
 
Again the nearby system, Arp 220, provides more details, as
well as unanswered questions.  BeppoSAX observations \citep{iwa01} placed
severe constraints on an energetically significant AGN in Arp 220,
requiring an absorbing column, $N_H > 10^{25}$ cm$^{-2}$ to hide an X-ray AGN.
They suggested X-ray binaries as the source of Arp 220's hard
X-rays. \citet{cle02} detected several sources near Arp 220's nucleus,
including a mildly absorbed point source with a hard spectrum coincident
with the western radio nucleus, plus a fainter source which may coincide
with the eastern nucleus.  A classical X-ray AGN cannot be ruled out, but
again, columns greater that $5 \times 10^{24}$ cm$^{-2}$ would be required to
hide it. \citet{iwa} have reported the detection of Fe K emission in the
XMM-Newton spectrum of Arp 220.  A supernova shocked bubble as suggested by
the VLBI observations of radio supernovae could provide an explanation.
However, the apparent lack of emission from X-ray binaries is incompatible
with the high supernova rate ($\sim$2 SNe yr$^{-1}$) required.

\section{The emerging picture of ULIRGs:  Local Universe}\label{sec:picture}

Taken as a whole, the observations of local Luminous Infrared Galaxies
suggest that, at lower luminosities at least, they are dominated by
starburst emission.  The continuity of the $L_{HCN}\, vs\, L_{FIR}$ relation
\citep{gs04} provides circumstantial evidence that most, if not all,
Luminous IR Galaxies are starburst powered.  Their sample, however, as well
as many others quoted above, suffers from the low space density of ULIRGs,
with only 6 classical ULIRGs ($L_{IR} > 10^{12}\, L_\odot$) and only two
with $L_{IR} > 10^{12.2}\, L_\odot$).  The LRSN discovered in compact
ULIRGs would seem to indicate starburst dominance in these systems, however
we do not understand enough about these LRSN to establish a star-formation
rate or a starburst bolometric luminosity, leaving open the possibility of
significant if not dominant AGN power even in the best studied source Arp
220.  ULIRGs are generally under-luminous in the X-rays compared to
classical AGN, requiring sensitive, high energy, observations to detect
them.  Evidence for very high X-ray obscuring columns is mounting, which
could explain the low observed flux levels, and estimates of the importance
of the X-ray-detected, obscuration-corrected AGN power to ULIRG energetics
range from minor to dominant.  Spectroscopy from the optical through
mid-infrared suggests that the incidence of AGN and their strength relative
to the ubiquitous starbursts increases with FIR Luminosity and that AGN may
dominate for $L_{IR} > 10^{12.5}\, L_\odot$, but there are very few such
ULIRGs in most local samples.  Mid-infrared high excitation diagnostic
lines are a very promising line of investigation, however very deep
silicate and ice absorptions are also detected in some ULIRGs, so even
these mid-IR diagnostics may severely under-estimate AGN energetic
contributions due to extinction.

And so the debate continues, and the very fact that it continues two
decades after IRAS highlights that most ULIRGs almost certainly contain
both a starburst and a monster, and that the key question really concerns
the connection between the two, be it evolutionary (one evolves into the
other), causal (one triggers the other somehow) or coincidental (another
influence triggers both).  We cannot address the wealth of literature on
this topic here, involving as it does evolutionary schemes of all kinds and
their implications, and extensive comparative AGN and starburst population
studies and demographics, and so we outline here only a few key points.
One popular scenario is an evolutionary sequence in which major gas rich
galaxy mergers first result in a massive cool starburst-dominated ULIRG,
followed by a warm ULIRG as a QSO turns on inside the dust cocoon and heats
the surrounding dust, and then finally the QSO emerges in an optically
bright phase when it blows away the surrounding dust cocoon, and the
resulting stellar system resembles a spheroid (Sanders et al. 1988a;
Kormendy and Sanders 1992; Joseph 1999; Fabian 1999; Lipari et al. 2003).
Another important scenario is unification-by-orientation of AGN, in which a
broad-line (type I) active nucleus is highly obscured by a dusty molecular
torus (or other non-symmetrical geometry) when viewed off-axis, so that an
optically bright QSO viewed off-axis would appear as a ULIRG.  Both of
these scenarios predict close relationships between starbursts and AGN.

The evolutionary scenarios have recently received a boost from a sequence
of papers reporting high resolution hydrodynamic simulations of gas rich
major mergers (Di Matteo et al., 2005; Springel et al., 2005; Hopkins et al.,
2005a,b,c,d), motivated by linking the growth of spheroid masses and
supermassive black hole (SMBH) masses in order to explain the the observed
correlations between SMBH mass and bulge mass (Magorrian et al. 1998) or
velocity dispersion (Ferrarese and Merritt 2000, Gebhardt et al. 2000) of
local spheroids.  Accretion rates are predicted to be highest at late
merger stages, when the SMBH grows exponentially, followed by the most
luminous optically-visible QSO phase when the active QSO essentially
explosively drives out all remaining material in the system (Hopkins et
al. 2005a).  The period of high obscuration during the high accretion rate
phase would correspond to an obscured QSO, ie an AGN-powered ULIRG.
Starburst events occur earlier in the lifetime of the merger when gas is
still plentiful, and a starburst-ULIRG phase could occur when the gas is
centrally concentrated into a dense compact region at relatively late
stages.
 
Hopkins et al. (2005e) have made approximate predictions of ULIRG space
densities based on these simulations, estimating at z=0.15,
3$\times$10$^{-7}$ and 9$\times$10$^{-8}$ Mpc$^{-3}$ at infrared
luminosities of 1.6 and 2.5 $\times$ 10$^{12} L_{\odot}$, respectively,
which is in good agreement with the 1Jy survey of Kim and Sanders (1998).
Hopkins et al. also predict a decrease by a factor of 1.5 to the lower
redshift, z=0.04, which also agrees well with the Kim and Sanders work, and
an increase to z$\sim$1-3 of ${\Phi}(L>10^{11}L_{\odot})$ of 1-3 $\times$
10$^{-5}$ Mpc$^{-3}$, which they compare to the sub-mm number counts of
Barger et al. (2005).  \citet{cha05} have derived the first estimate of
ULIRG luminosity functions in this redshift range, though sensitive only to
systems above about 10$^{12.3} L_{\odot}$.  They find space densities
$>$6 $\times10^{-6}$ Mpc$^{-3}$ in their luminosity range, which could be
significantly higher than the Hopkins et al. predictions, depending on the
unmeasured break of the IR ULIRG luminosity function at these redshifts.
Deep Spitzer surveys should help extend the ULIRG luminosity function to
lower luminosities in this redshift range.
 
These ULIRG-QSO evolutionary sequences are appealing and seem likely to be
correct for at least a fraction of the ULIRG population.  Many predictions
remain to be verified, however, and alternative scenarios are also popular.
A key test is the host morphology of ULIRGs and optically-bright QSOs.  As
we have described in Section 3, not all ULIRGs are found at late stages of
a merger, although most are indeed found to be associated with the merger
process.  Farrah et al. (2001) have proposed an alternative
merger-evolution scheme in which the stage at which a ULIRG occurs depends
chiefly on the morphological type of the two interacting systems.  Deep HST
imaging of QSO hosts is now revealing that significant numbers also appear
to inhabit disturbed systems \citep{limh,per01,sanc}, supporting an
evolutionary connection between starbursts and QSOs, although some QSOs
appear to occur in spiral hosts with very little evidence of recent
disturbance.  This latter result has been taken to indicate that they could
not have been triggered by a recent major merger, however recent
simulations have demonstrated that major gas-rich mergers can result in
disk systems under certain circumstances \citep{spri05,rob05}

Genzel et al. (2001) and Tacconi et al. (2002) observed NIR structural 
properties and stellar dynamics of 18 z$<$0.18 ULIRGs, showing that 
ULIRG hosts lie on the fundamental plane
of elliptical galaxies and thus are very likely to evolve into ellipticals,
but that they have significantly lower mass hosts than a luminosity-matched
sample of radio-loud and radio-quiet QSOs from Dunlop et al. (2003). 
A difficulty with this approach is that both starbursts and
QSOs are expected to have rapidly varying luminosity during the merger
sequence, so it is not clear that samples should be luminosity
matched to compare their host masses.   These authors 
also reviewed the cluster environment of local ULIRGs, QSOs (from the study
of McLure \& Dunlop 2002) and ellipticals, finding that while the
ellipticals and quasars are found in all environments, none of the 117
ULIRGs that were investigated are located in an environment richer than a
small group, which provides some statistical evidence that at least those
quasars found locally in rich environments may not have evolved via a ULIRG
phase.  

Haas et al. (2003) searched for a red transition population between
obscured and unobscured QSO populations amongst Palomar-Green (PG) QSOs.
Optically-selected QSOs as a class have a very high incidence rate of
luminous FIR and submm emission, consistent with most of them qualifying as
ULIRGs by IR luminosity level (Haas et al., 2003; Polletta et al., 2000;
Hatziminaoglou et al., 2005).  Haas et al. suggested a sequence of near-,
mid- and far-infrared SEDs for their PG QSO sample which could plausibly
represent the transition from a young dust-obscured QSO through the stages
of re-distribution of the dust and settling of the dust into a torus, and
which is less well explained by simple orientation effects alone.  Late
evolution-stage systems may be expected to have low fuel supplies, perhaps
as seen some moderate redshift HLIRGs with low CO masses and
$L_{IR}/M_{gas}$ ratios or which lack cool dust components ({\it eg.} Yun
\& Scoville 1998; Verma et al., 2002). Another suggested young QSO transitory
class are the broad absorption line (BAL) QSOs, especially those with
strong FeII and weak [OIII] emission, which are preferentially found to be
infrared luminous and which could therefore be young dusty QSOs (Voit et
al., 1993; Lipari, 1994; Egami et al., 1996; Canalizo \& Stockton 2001,
Lipari et al., 2003).  FeLoBAL QSOs could alternatively
be older QSOs viewed preferentially along the radial surface of the torus
(Elvis 2000).
 
Another class of potentially young broad-line QSOs are the red 2MASS QSOs,
selected to have J-K$>$2 and expected to be significantly dust-obscured
(Cutri et al. 2002).  These objects tend to be type-1 AGN with
moderate luminosities and z$<$0.8, the relatively low luminosities and
redshifts being due to a selection bias due to the near-infrared
k-correction. \citet{leip} extended this red 2MASS QSO search
using ISO 6.7$\mu$m data, and found a surface density of 1.5 times that of
SDSS QSOs to the same optical magnitude depths, and also that the 2MASS-ISO
QSOs are significantly redder than SDSS QSOs, and that SDSS QSO colour
selection criteria would have missed about 1/3 of these red QSOs.

Most detailed imaging and morphological studies so far have focused on
relatively low redshift systems.  At higher redshifts we might expect
different triggering mechanisms under different environmental
circumstances, which we discuss further in Section 5.
 
\section{ULIRGs at high redshift}\label{ulirghigh}

\subsection{The cosmic infrared background}

The existence of a significant population of IR-luminous galaxies at high
redshift, though hinted at in followup of IRAS, was thrown to the fore of
astronomical debate by results from the Cosmic Background Explorer (COBE)
satellite. Launched on November 18, 1989, COBE was responsible for two of
the most significant astronomical results in the 20th century. The first
was measuring the spectral shape and level of anisotropy in the Cosmic
Microwave Background (CMB, peaking at $\sim2000\mu$m), providing
overwhelming evidence for the Hot Big Bang model and giving a glimpse at
the early structures that would eventually evolve into todays galaxies and
clusters. The second result, more relevant for this review, was the
discovery of a Cosmic Infrared Background (CIB) with FIRAS at 240$\mu$m
\citep{pug}, and at 140$\mu$m and 240$\mu$m with DIRBE \citep{schl,hau,fix}
(later detections with DIRBE of the CIB at 2.4$\mu$m and 3.5$\mu$m have
been published, see Hauser \& Dwek, 2001; and Kashlinsky, 2005 for
reviews). Such an infrared background had been predicted many years
previously \citep{par}, but had proven fiendishly difficult to detect,
remaining invisible both to rocket-borne IR observatories \citep{kaw}, and
to IRAS \citep{rr90,oli92}. COBE was the first observatory with both the
right instrumentation and sufficient sensitivity to detect the CIB.

The cosmological implications from the discovery of the CIB were profound
\citep{dwe}.  The total background detected by COBE between 140 and 5000
$\mu$m is $\sim16$ nW m$^{-2}$ sr$^{-1}$, or 20\%-50\% of the total
background light expected from energy release by nucleosynthesis over the
entire history of the Universe, implying that ~5\%-15\% of all baryons are
or have been parts of stars. And while the CIB itself amounts to less than
about 2\% of the CMB, the intensity of the CIB is still surprisingly high,
comparable to, or exceeding, the integrated optical light from the galaxies
in the Hubble Deep Field \citep{hau}.

When
compared to the cosmic history of star formation derived from optical and
UV surveys \citep{mad} a serious discrepancy became apparent; the
CIB detected by COBE requires at least a factor of two more star formation
than was apparent in optical and UV surveys, meaning that the integrated
star formation rate at $z\sim1.5$ must be higher than that implied from
UV/optical observations by a comparable factor, and that this star
formation must be largely surrounded by dust. COBE however could provide
little further constraint on the form of the star formation history implied
by the CIB; in principle there could be large numbers of galaxies at high
redshift that are faint in the IR, or a few extremely IR-luminous
sources\footnote{Though even at this stage source count models that could
explain the CIB all invoked a population of high-redshift IR-luminous
starbursts \citep{gui,bla2,rr01,xu01}}. Further progress required that the CIB
be resolved into its constituent sources.
 
\subsection{Resolving the CIB: LIRGs and ULIRGs at $0<z<1.5$}

The first major steps in resolving the CIB came from extragalactic surveys
carried out by ISO at 7$\mu$m and 15$\mu$m with ISOCAM, and at 90$\mu$m and
170$\mu$m with ISOPHOT, most notably observations of the HDF
\citep{oli97,rr97}, the European Large Area ISO Survey (ELAIS;
Oliver et al., 2000; Rowan-Robinson et al., 2004), and the FIRBACK 
survey \citep{pug2,dol} (see also reviews in Elbaz, 2005; Verma et al., 2005;
Oliver \& Pozzi, 2005).  The 15$\mu$m surveys
\citep{elb} were particularly successful; the sources seen in these surveys
could, with considerable extrapolation of the SED shape at longer
wavelengths, account for around 80\% of the CIB. Followup observations
showed these sources have $<z> {\sim} 0.8$, and that the comoving density
of infrared light due to these 15$\mu$m sources is at least 40 times
greater at $z\sim1$ than in the local Universe (compare this to the B band
luminosity density, which is only about three times the local value at
$z\sim1$). Not to be outdone, the FIRBACK survey found that the 170$\mu$m
source counts show very strong evolution with redshift (reaching z$\sim$1),
directly resolve 5\% of the CIB \citep{dol}, and are responsible for a
dramatic rise in the integrated star formation rate, with a value at least
ten times that seen locally at $z\sim1$ \citep{rr97,flo,poz}. Overall,
therefore, the ISO deep surveys put the IRAS discovery of strong IR galaxy
evolution onto a very firm footing. Out to z$\sim$1 the ISO sources are
chiefly LIRGs, not ULIRGs, and they are generally similar in many ways to
lower redshift LIRGs, although possibly with lower average metallicity
\citep{fra2,elb2,lia}.

The 170$\mu$m surveys also resulted in another interesting discovery; a
large number of the 170$\mu$m-selected objects have cooler dust
temperatures and larger dust masses than those seen in starburst galaxies
selected at mid-IR wavelengths. This is not unexpected for a long
wavelength-selected survey, since a large luminosity from a cool dust
component requires a larger dust mass than a warmer source. ISO however was
the first observatory to show conclusively that large masses of 'cool' dust
existed in many galaxies, from local spirals to distant ULIRGs. In local
spirals this cold ``cirrus'' component is expected to be diffuse dust
heated by the interstellar radiation field from later type stars.  In
ULIRGs however this cold component could alternatively be a compact dusty
starburst with colder than average dust. This discovery has interesting
implications for studies of ULIRGs at higher redshifts, as we will discuss
in Section \ref{ulirgsubmm}.

Focusing on ULIRGs; the largest ISO survey, and therefore the one most
likely to discover ULIRGs in any number, was ELAIS (see Fig. 1). The most
sensitive ELAIS band for LIRGs and ULIRGs was 15$\mu$m, resulting in the
detection of just under 100 ULIRGs, comprising $>$10\% of the 15$\mu$m
sample. Around 10 of these are HLIRGs; the first of these to be identified
being a QSO at z=1.01 \citep{mor}. These ULIRGs range in redshift up to
$z\gtrsim3$ and many are fit well in colour by an Arp 220-like SED. The
longer wavelength ELAIS surveys result in many fewer sources due to
decreased sensitivities; \citep{tay} have identified 4 likely ULIRGs in the
FIRBACK-ELAIS N2 (FN2) 170$\mu$m population, while \citep{den} identify 1
or 2 in the FIRBACK-ELAIS N1 (FN1) field. \citep{saj} identified a
population of z$\sim$0.5-1 ULIRGs, representing $\sim$ 1/6 of the total
170$\mu$m FN1 sample, and \citet{cha02} identified two of these at z=0.5
and 0.9 to be unusually cold systems with merger morphologies.  Several
sources in each field remained optically unidentified and are likely to be
ULIRGs at moderate redshifts (0.5-1). Several ULIRGs and HLIRGs have been
discovered in other ISO surveys, including a z=1.7 FeLoBAL QSO behind the
z=0.56 cluster J1888.16CL \citep{duc02}, and a radio detected ERO at z=1.5
\citep{pie}, probably a QSO with very hot dust.

Follow-up of $z\gtrsim0.5$ ULIRGs revealed that many are involved in
ongoing interactions, but that interestingly a significant number are QSOs
with very massive ($>2L^{*}$) elliptical host galaxies
\citep{far3}. Detailed SED modeling \citep{rrc,rr2000,verma02,far4} showed
evidence for both starburst and AGN activity in most sources, with colossal
implied star formation rates of up to $\sim1000 M_{\odot}$
yr$^{-1}$. X-ray observations however found unexpectedly weak X-ray
emission, implying either no obscured AGN or extremely high obscuration
levels \citep{wil0}; the latter interpretation is favored by deeper
observations which found evidence for Compton thick AGN in some (but not
all) of these systems \citep{wil1}.  Very recently, \citet{iwa} have used
XMM to detect faint X-ray emission with a hint of the 6.4 keV Fe K$\alpha$
line from F15307+3252, which they attribute to a Compton thick AGN with
$L$(2-10kev)$>10^{45}$ erg/s, which can account for a significant fraction of
the infrared luminosity.

The picture that emerged from ISO, therefore, was one where very IR-luminous
galaxies become substantially more numerous with increasing redshift,
making $z\sim1$ LIRGs and ULIRGs a cosmologically significant
population.  The $0.5<z<1.0$ LIRG population seems quite similar to
LIRGs at lower redshift, but also revealed a population of moderate redshift 
LIRGs with relatively cool dust emission. The higher redshift ULIRG 
population exhibits a more prevalent level of QSO activity than found in their 
local cousins, a result that could be either luminosity or redshift related 
since the most luminous ULIRGs are found at the higher redshifts.   

\subsection{Resolving the CIB: ULIRGs at $z>1.5$}\label{ulirgsubmm}

The CIB had another startling surprise in store. Though ISO had proven
itself remarkably adept at resolving much of the CIB, there remained a
significant shortfall between the CIB as measured by COBE, and the sources
detected by ISO. The background levels at ${\lambda}>200{\mu}$m determined by 
FIRAS onboard COBE \citep{fix} implied a population of ``colder'' sources, 
probably $z\gtrsim1$ systems with IR emission redshifted to longer wavelengths 
than the sources detectable in shallower, ${\lambda}<200{\mu}$m, ISO surveys. 

Fortunately, the perfect instrument to search for just this kind of source
had just been commissioned at the James Clerk Maxwell Telescope, namely the
Sub-millimetre Common User Bolometer Array (SCUBA, Holland et al. 1999). SCUBA
heralded the advent of bolometer arrays for sub-mm observations, and was
one of the first sub-mm instruments that could map large areas of sky
relatively quickly in the sub-mm; operating at substantially longer
wavelengths than ISO, namely 850$\mu$m and 450$\mu$m, SCUBA had the
sensitivity to detect these distant `cold' ULIRGs. Furthermore, the
strongly negative k-correction at sub-mm wavelengths \citep{blal} meant
that a ULIRG at $z=5$ is (depending on the choice of cosmology) 
almost as easy to detect as a ULIRG of comparable luminosity at
$z=1$. Other important sub-mm bolometer arrays include the 350$\mu$m
optimized SHARC-II camera and the 1.1mm Bolocam instrument, both at the
Caltech Sub-mm Observatory, and the MAx-Planck Millimetre Bolometer (MAMBO)
at the Institut de Radioastronomie Millimetrique (IRAM) 30m telescope,
which operates at 1.2mm.

The advent of sub-mm array instruments therefore prompted a plethora of surveys
try to find these distant ``cold'' sources, ranging from ultra-deep
surveys of lensing cluster fields \citep{sma97,sma1,ivi0} through deep
blank-field surveys of small areas \citep{hug,barg}, to wider area
blank-field surveys to shallower depths \citep{eal,scot,bor,gre0,lau}. 
These surveys found a huge population of sub-mm bright, optically faint
sources, with source counts of $320^{+80}_{-100}$ and $180\pm 60$ per
square degree at $S_{850}>10$mJy \citep{scot}.  Taken together, the sources
found in the various sub-mm \& mm surveys can directly account for around 50\%
of the CIB detected by FIRAS, and with reasonable extrapolation, can
account for all of it \citep{barg3,bla3}.

The first step was to determine whether these sub-mm sources really are
distant ULIRGs, or some other class of object.  If, as is very plausible,
the sub-mm emission from distant galaxies can be modeled with a simple
modified blackbody, where the dust temperature distribution is
characterised by $T$, then the flux at frequency $\nu$, $F(\nu)$, is given by
$F(\nu)\propto\nu^{\beta}B(\nu,T)$, where $\beta$ is the spectral index
(also referred to as the emissivity), and $B(\nu,T)$ is the Planck
function. As a virtue of the shape of this function, sources with a
redshift of $z\gtrsim 1$, a dust temperature of $\gtrsim25$K, an 850$\mu$m
flux of $S_{850}>1$mJy, and a reasonable choice of emissivity will, under
most circumstances, be ULIRGs. Simple photometric redshift estimates
\citep{ber,fox,bor2,web} place virtually all the sources from sub-mm
surveys at $z>1$, and most at $z>2$, meaning that the sub-mm sources are
likely to be extremely luminous ULIRGs, with star formation rates
substantially exceeding $1000$M$_{\odot}$ yr$^{-1}$. This implies that
there are of order a few hundred ULIRGs per square degree at $z>1$,
compared to about one ULIRG every four square degrees locally; though
contributing less than 1\% of the total extragalactic background light
locally, ULIRGs were therefore shown to be a major, perhaps even dominant
contributor at $z>1$.

Before proceeding further, there is an important cautionary note. A single
sub-mm flux cannot be used to infer dust mass or source luminosity without
assuming a dust grain size and temperature distribution \citep{hil}, thus,
for example, large cold disks can be confused with compact warmer
starbursts of very different bolometric luminosity
\citep{kav,ef3,far6,alm05}.  If the redshift is also unknown, a much lower
luminosity foreground cold disk or even a Galactic dust cloud could easily
mimic a high-redshift ULIRG \citep{law}.  As we describe below, however,
there is now good evidence that at least $\sim$70\% of sub-mm sources are
indeed $z>1$ ULIRGs, but within the remaining 30\% there remains the
possibility of significant sample contamination.  Moreover, a strong
temperature selection bias does exist in these very long-wavelength
surveys; for example, a ULIRG at $2<z<3$ with $L_{ir}\sim10^{13}L_{\odot}$
and a dust temperature of 60K would not be detected in the wide, shallow
surveys with SCUBA \citep{bla5}. Such ``hot'' sources do appear to exist
\citep{cha042,lut}, and could be responsible for up to 1/3 of the DIRBE
detected CIB. Therefore, sub-mm surveys alone do not necessarily provide a
hyaline view of dust-shrouded star formation, or even of ULIRG activity, at
high redshift. Spitzer is proving to be extremely important for discovering
the warmest high-redshift ULIRGs (see Figure 1), as we shall describe
further below.

\begin{figure}
\centering
\includegraphics[height=8cm]{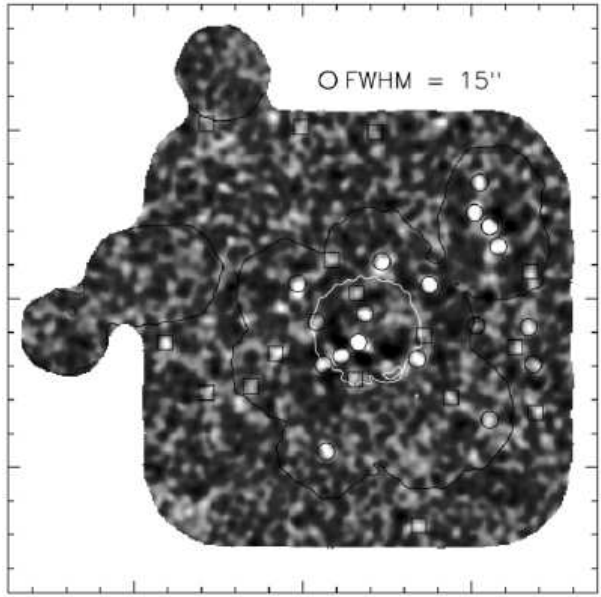}
\caption{An example of a recent submm survey; in this case the HDF-N SCUBA 
``Supermap'' \citep{bor}. circles = $> 4\sigma$, squares $= 3.5-4D\sigma$. 
The white line encompasses the \citet{hug} SCUBA-HDF observations.}
\label{submmmap}       
\end{figure}

Due to the coarse angular resolution of SCUBA (and indeed most sub-mm \& mm
observatories up to now) and the optical faintness of most of these
sources, determining reliable ID's for follow-up proved to be extremely
difficult (see Dunlop et al. 2004, for a perfect example). Nevertheless, since
1997, progress has slowly but surely been made in following up Sub-mm
Galaxies (SMGs), and a clear picture of their nature has now started to
emerge.   A prime question is: how well do these distant systems resemble local
ULIRGs?  We would not necessarily expect them to be closely similar in
nature because conditions are very different at high redshift, notably
an earlier stage in the clustering evolution of the underlying dark matter, 
and larger gas fractions.

Optical and near-IR imaging \citep{sma0,web,sma2,pop} has shown that SMGs
have a diverse range in optical properties, ranging from optically bright
sources, to sources undetected in even the deepest optical imaging. Radio
observations, particularly with the VLA, provided a major step forward in
followup of SMGs \citep{ivi}, detecting around 70\% of the SMGs found in
wide-field surveys with sufficiently high spatial resolution to allow
reliable identification of optical/near-IR counterparts. Later, very high
resolution radio studies have shown, intriguingly, that the radio emission
in a surprisingly high fraction of SMGs, around 70\%, is spatially extended
on scales of $\sim10$ kpc \citep{cha04}, significantly larger than the dust
emission seen in most local ULIRGs, which are generally less than a kpc
across \citep{dow}.  The beam shape, especially when beam smearing is
accounted for, is probably not well enough known to provide much evidence
on the intrinsic shape of these radio sources, so possible explanations
include large scale star formation over large disks, multiple compact starburst
sites in a large disk or a coalescing group, or to AGN jet-induced star
formation as seen in some distant radio galaxies \citep{vanB}. Another class
of z$>$2 large scale systems inferred to have ULIRG-level luminosities 
are the so-called Ly$\alpha$ blobs (Geach et al., 2005).    

Early efforts to obtain spectroscopic redshifts for SMGs had been mostly
stymied (but not entirely, see Barger et al., 1999a), not only by uncertain IDs
and extreme optical faintness, but also because many SMGs lie within the
so-called redshift `desert', $1<z<2$, within which the bandpasses of most
optical spectrographs could only sample weak absorption features, from
which redshifts are hard to determine. But with high resolution radio ID's
in hand, and significant tenacity, spectroscopic redshifts for SMGs were
soon forthcoming \citep{cha05,sim}. These surveys showed that SMGs have a
median redshift of 2.4, removing any last lingering
doubts that the (radio detected) SMGs were indeed high-redshift ULIRGs. The
optical spectra themselves show evidence for both starbursts and AGN in
most sources, high unobscured star formation rates, and that SMGs are
generally metal rich \citep{swi}. Very recent optical spectroscopic studies
have focused on integral field spectroscopy, though instrumentation is only
now becoming capable of integral field studies of distant SMGs. Early
results however look very promising; the host galaxies of SMGs are metal
rich, and appear to contain surprisingly large numbers of evolved stars and
show evidence for ongoing interactions \citep{tec,swi2}.

With accurate redshifts it was possible to use heterodyne instruments
(which generally have a very narrow bandpass, making spectroscopic
redshifts essential for meaningful observations) to look at far-IR fine
structure lines, particularly those of CO. CO surveys of SMGs
\citep{gen2,ner,gre} revealed that they have enormous masses (often
exceeding $\sim1\times10^{11}M_{\odot}$) of molecular gas, usually
unresolved but in some cases extended over 3-5kpc \citep{gen2}, much
greater than those masses seen in local ULIRGs and comparable to those seen
in high-redshift radio galaxies. HST imaging shows that SMGs often possess
disturbed morphologies, consistent with (but not solely supportive of)
ongoing major mergers, similar to the local ULIRG population
\citep{sma,cons}.  Very recent results based on deep X-ray observations
have provided convincing evidence that the bulk of SMGs harbour a Compton
thick AGN in addition to a starburst, implying that a sub-mm galaxy
signposts a major period of growth of the central supermassive black hole
in these systems, as well as an intense starburst \citep{ale}. There is
some evidence that the periods of black hole growth and stellar mass
buildup in these systems may not be coeval \citep{bor5}, although there are
uncertainties in the derivation of the X-ray obscuring columns, and
therefore of the black hole mass \citep{poll06}.

Further results have come from the first observations of SMGs with
Spitzer. Results from IRAC and 24$\mu$m imaging \citep{ega,fray} have shown
that SMGs have a wide range of observed mid-IR colours, and that their
rest-frame mid-IR SEDs can in most cases be fitted with a starburst SED
template, with the remainder being well fitted by a power-law AGN-dominated
template. Photometric redshifts estimated from the near-infrared hump
peaking at rest-frame 1.6$\mu$m and detected in the Spitzer IRAC bands
(3.6-8.0$\mu$m) span $1<z<3.5$. \citet{ivi2} observed 9 MAMBO
1200$\mu$m sources in the same region, finding 75\% of them to be
starburst-dominated using Spitzer mid-infrared spectral shapes (assuming
redshifts $\sim$2.5), and concluding that the more AGN-like SEDs are
consistent with AGN indications in these objects from UV/optical
spectroscopy and X-ray imaging.  

Spitzer will be a very powerful tool for finding large populations of
ULIRGs at $z>1.5$, with many hundreds of candidates appearing in the
Spitzer surveys such as SWIRE (Lonsdale et al. 2006), the Bootes
Shallow field, the First Look Survey, the GTO deep
surveys and GOODS (see Figure 1). Moreover Spitzer complements the cold
dust selection function of the sub-mm and mm surveys with its mid-IR selection
function and enhanced sensitivity to warm systems.  

The first source counts and luminosity functions at 24$\mu$m, 70$\mu$m and
160$\mu$m from Spitzer extend the results seen by ISO.  Spitzer is
revealing large numbers of systems with significantly more extreme
infrared/optical luminosity ratio than seen in the local Universe
(Rowan-Robinson et al. 2005), a result also known from moderate redshift
ISO surveys (Rowan-Robinson et al. 2004; Oliver \& Pozzi 2005).
Rowan-Robinson et al. (2005) also find a significant population of very
cool luminous Spitzer systems, which could be very large disks with
quiescent star formation rates rather than starbursts.  The 24$\mu$m counts
in the GOODS fields \citep{pap} show evidence for strong evolution,
exceeding the `no evolution' counts predictions by a factor of at least 10,
and implying evolution in the comoving IR energy density of the form
$(1+z)^{3.9\pm0.4}$ up to $z\sim1$ \citep{lef}. The Spitzer counts at
70$\mu$m and 160$\mu$m \citep{dol2} directly resolve 20\% and 7\% of the
CIB at these wavelengths, respectively, and also show strong evolution, by
a factor of $\sim3$ over no evolution models, implying that the galaxies
responsible for this background mostly lie in the redshift range
$0.7<z<0.9$. \citet{perz} used find that LIRGs and ULIRGs are increasingly
important contributors to the infrared energy density as redshift increases
to z$\sim$3, being responsible for half of all star formation by
z$\sim$1.5. The characteristic luminosity of the luminosity function,
$L^*$, increases steadily with z, consistent with the cosmic star formation
rate density going as $(1+z)^4$ to z$=0.8$, then flattening somewhat, and
with ULIRGs playing a rapidly increasing role above z=1.3. \citet{lef} came
to similar conclusions out to z$\sim$1, and also compared the UV and IR
star formation history since z=1 using deep observations from Spitzer and
GALEX, finding that the SFH has evolved much more strongly in the IR than
in the UV, thus confirming the ISO-based results of \citet{rr97,flo,poz}.

Yan et al. (2005) and Houck et al. (2005) have reported the first Spitzer
low resolution IRS spectroscopy for extreme IR/optical z$\sim$2 ULIRG
candidates, demonstrating that Spitzer is capable of determining
mid-infrared redshifts for the brightest mid-infrared galaxies
($f_{24}>0.75$mJy).  The space density of $L_{ir}>10^{12.3}L_{\odot}$
z$\sim$2 Spitzer mid-infrared-selected ULIRGs may be similar to that of
SMGs (Yan et al.) at this redshift, though, the sample sizes are as yet
small and incomplete.  Yan et al. infer a higher fraction of
starburst-dominated ULIRGs in their sample than Houck et al., which is
probably due to the additional selection criterion on high 24/8$\mu$m colour
({\it ie} a rejection of mid-IR warm sources).  Lutz et
al. (2006) report a low 1.2mm detection rate these systems using MAMBO, and
suggest that their mid-infrared-selected sample may have significantly
warmer dust than submm-selected samples.  Franceschini et al. (2004),
Martinez-Sansigre et al. (2005), Donley et al. (2005) and Polletta et
al. (2006) have used Spitzer surveys to discover extensive populations of
heavily obscured AGN populations that can have very high X-ray obscuring
columns, including two z$>$2, Compton-thick SWIRE HLIRGs with AGN
torus-dominated mid-IR SEDs \citep{poll06}.  These new Spitzer results
indicate that in many ULIRGs highly obscured QSOs are extremely difficult
to detect at any wavelength, including hard X-ray and mid-infrared, due to
extreme optical depths.

Daddi et al. (2005) have reported the first Spitzer results from the Great
Observatories Origins Deep Survey (GOODS) for z$\sim$2 ULIRGs, finding that
typical massive ($M_{stellar}{\sim}10^{11}M_{\odot}$) star forming galaxies
at this redshift are ULIRGs, based on 24$\mu$m and radio detections, with a
co-moving space density of ULIRGs at z=2 at least 3 orders of magnitude
greater than the local one.

In summary it would be fair to say that surveys with SCUBA and other sub-mm/mm
observatories, and their followup, have revolutionized our perceptions of
ULIRGs, while Spitzer stands poised to contribute its own revolution of
knowledge about them.  From being little more than an interesting oddity in
the local Universe, ULIRGs have become a crucially important population at
$z>1$. Furthermore, whilst $z>1$ ULIRGs are similar in many ways to their
local counterparts - both populations apparently being heavily
dust-obscured starbursts and/or AGN triggered by interactions between
large, evolved systems - there are important differences.  High-redshift
ULIRGs are substantially more gas rich than their local counterparts, and
some may have extended rather than compact starbursts. 
Moreover the star formation rates implied (if AGN do not strongly
dominate the energetics) are enormous, pushing the limits on theoretical 
ideas for a ``maximal'' starburst in a massive galaxy: 
M$_{gas}/t_{dyn}$ = 10$^{11}$ M$_{\odot}$ / 10$^8$ yrs $\sim$1000
M$_{\odot}/$yr \citep{egg62}.    To understand
these differences, and place these high-redshift ULIRGs into a cosmological
context, we must turn to theories of galaxy and large-scale structure
formation.

\section{ULIRGs and Large-scale structure}\label{ulirglss} 
The observation
that these distant ULIRGs appear to be very massive systems means that
large amounts of raw material are needed to make one, implying that ULIRGs
are likely to be found in regions with large concentrations of
baryons. Therefore, in order to understand when, where and why ULIRGs exist,
we must turn to galaxy and large-scale structure formation models to see
where such large concentrations of baryons are predicted to be
found. Modern structure formation models describe the evolution of the
total mass distribution (comprising both non-baryonic and baryonic `cold'
dark matter) via the evolution of initially Gaussian density fluctuations,
which means that the evolution of the total mass distribution should be
traced by the formation and evolution of galaxies in some determinable
way. Density fluctuations in the dark matter distribution, commonly
referred to as `halos', are predicted to undergo successive mergers over
time to build halos of steadily increasing mass. Galaxies are predicted to
form from the baryonic matter in these halos, driven either by inter- or
intra-halo mergers, or passive collapse of baryons to the halo central
regions, or some combination of all three mechanisms \citep{mez,kob}. This
framework of `biased' hierarchical buildup \citep{col,gran,hatt}, coupled
with a $\Lambda$ cosmology \citep{spe}, has proven to be remarkably
successful in explaining many aspects of galaxy and large-scale structure
formation.

Within this framework, the largest concentrations of baryons are expected
to be found within the most massive dark matter halos. A basic
verification of this idea can be seen in the local Universe, where the most
massive elliptical galaxies are almost invariably found within very rich
galaxy clusters; which presumably correspond to large overdensities of
baryons in the local Universe. The formation history of these very massive
galaxies, even before the advent of sub-mm surveys, was controversial.  The
`naive' expectation from hierarchical theory would be that massive galaxies
form relatively late and over a long period of time, representing
successive mergers between dark matter halos, building up the required
large baryon reservoirs \citep{bau0}, and indeed some massive galaxies do
appear to form in this way \citep{van,tam,bel}. There is however
observational evidence that many massive galaxies may form at high redshift
and on short timescales, for example from the discovery of evolved systems
at high redshift \citep{dun,mar,blak}, the extremely tight K-band Hubble
relation for radio galaxies \citep{deb2} implying a very rapid early
formation of the most massive galaxies \citep{roc}, and an inferred early
star formation epoch from the colour-magnitude diagram in clusters at
$z=0.5$ \citep{ell}, implying that the stars in very massive local galaxies
formed within a few Gyr of each other at $z>1$.

The SMGs and Spitzer high z ULIRG samples appear to be perfect candidates
for the formation events for those local massive ellipticals that appeared
to form very early and very fast; they lie at $z\simeq2.5$
\citep{sim,cha05}, which is a high enough redshift to make stars with the
ages seen locally, their star formation rates are high enough to make all
the stars in a local massive elliptical fast enough to satisfy the observed
colour-scaling relationships, and their comoving number density is
comparable to that of local $\gtrsim3L^{*}$ ellipticals
\citep{scot,cha042}. Serious theoretical difficulties remain however.
Early CDM-based galaxy formation models lacked the detailed astrophysics to
explain the number of observed SMGs and their star formation rates
\citep{gran,som,gui}.  Later models included improved treatments of
starbursts, but still had to include further modifications to explain the
observed sub-mm counts.  These include a top-heavy initial mass function
for the stars formed in bursts, so that the same sub-mm flux can be
produced with lower star formation rates \citep{bau} \footnote{an
interesting conundrum from this model is that while the sub-mm counts are
predicted to be dominated by bursts, these bursts would be responsible for
making only a small fraction of the stellar mass in evolved ellipticals in
the local Universe if their IMFs are top heavy.}, modified treatments of
virialization and the survival of subhalos to solve the `overmerging'
problem in the standard Press-Schechter formalism and make the SFRs high
enough \citep{vank,vank2}, and modifications to gas cooling and supernova
feedback \citep{gran2}.

Fundamentally, there are two testable basic predictions of these
hierarchical models for high-redshift ULIRGs. Firstly, these ULIRGs should
reside within rich environments, that is to say we should see clustering of
other galaxies around the ULIRGs on length scales of $<5$Mpc, representing
galaxy overdensities resulting from the local overdensity of
baryons. Testing this observationally is, however, difficult. At $z<1$
there have been numerous successful measurements of galaxy overdensities
around AGN using deep optical/near-IR imaging to find evolved systems
\citep{wol0,wol1}, but at the mean redshift of high-redshift ULIRGs
it is certainly not clear whether their local environments have built up
overdensities of evolved galaxies, even if they are overdense in
baryons. Secondly, we should observe the high-redshift ULIRGs themselves
to cluster together strongly on the sky, or in other words clustering of
just the high-redshift ULIRGs on length scales $\gtrsim25$Mpc. This
prediction is a consequence of the assumed form of the evolution of density
fluctuations with redshift; rare fluctuations in the underlying mass
distribution are predicted to cluster together on the sky, with the
strength of clustering depending on the degree of rarity \citep{bard}. Very
massive halos (those with masses $\gtrsim10^{14}M_{\odot}$), which are
universally predicted to be extremely rare at high redshift, are therefore
predicted to cluster together very strongly
(e.g. Moscardini et al. 1998; Kauffmann et al. 1999; Mo \& White 2002). 
This clustering should be traced by very
massive galaxies, as only the most massive halos contain sufficient baryon
reservoirs to form these galaxies. This also means that the clustering
strength of massive galaxies should translate straightforwardly to measure
of halo mass \citep{per}. Strong clustering among high-redshift ULIRGs is
also predicted if, as some suspect, they are part of a population uniting
QSOs and spheroids \citep{magl}. Measuring this form of clustering
observationally is however also difficult, requiring either $\gtrsim50$
sources with spectroscopic redshifts so that `associations' in redshift
space between different sources can be searched for, or a minimum of
several hundred sources with a reasonably well constrained redshift
distribution, and found over a large enough area of sky such that the
length scales of interest can be examined.

\begin{figure}
\centering
\includegraphics[height=12cm]{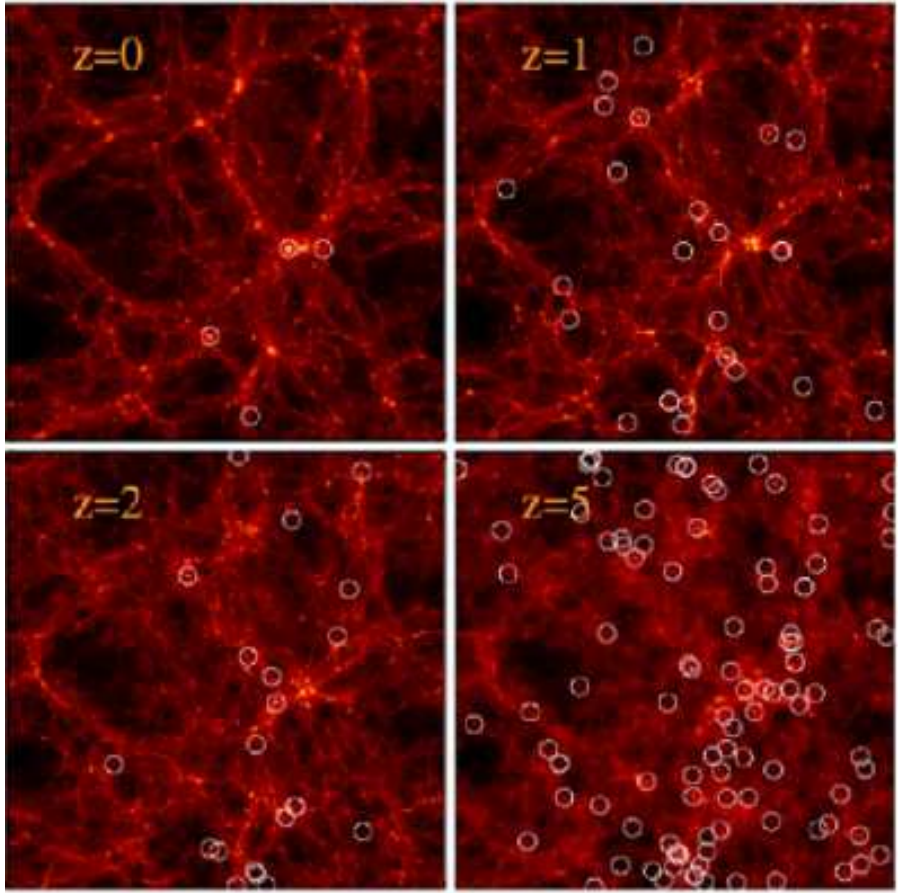}

\caption{An example of the cosmological hydrodynamic $\Lambda$CDM simulation (SPH G6 run)
described in Nagamine et al. (2005a,b).  Each panel has a comoving size of
143Mpc on a side, and the starforming galaxies with instantaneous SFR $>$
100 M$_{\odot}/$yr at each epoch are indicated by the circles (K. Nagamine, priv. comm.)}
\label{lssevo}       
\end{figure}

Despite these difficulties, numerous studies of both the environments of
high-redshift ULIRGs, and their clustering, have taken place, and while
these studies are not (yet) conclusive, they do generally lend support to
the idea that high-redshift ULIRGs form within massive dark matter
halos. Considering environment studies first; near-IR imaging
studies of the environmental richness of ULIRGs over $0.5<z<1.5$
\citep{far5} have shown that these systems reside in richer environments
than their local counterparts on average, and that some reside in moderate
rich clusters even at $z\gtrsim1$, though it is not clear whether this
represents a genuine increase in environmental richness for ULIRGs, or
whether this just reflects the global increase in `bias' over $0<z<1$. At
lower redshifts, LIRGs and ULIRGs have also been found in clusters
\citep{duc,lem,kle}. Turning to higher redshifts, then several authors have
noted overdensities of sub-mm sources around high-redshift systems that are
thought to reside within mass overdensities, including $z\sim4$ radio
galaxies \citep{ste,deb} (several of these sub-mm sources are also detected
in the X-ray, suggesting the presence of obscured AGN \citep{sma03} and
seen as $z\sim3$ Lyman Break galaxies \citep{cha0,gea}.

Turning to high-z ULIRG clustering; early efforts to measure angular
clustering on the sky came up against the limitations of available
instrumentation; at the time of writing there exists no sub-mm array
instrument that can map large enough areas of the sky to the required
depths (though SCUBA2, APEX and Herschel, are coming soon). No clear
picture has therefore emerged; some sub-mm surveys have uncovered tentative
hints of clustering on scales of a few arcminutes \citep{sco}, whereas
others show no signs of clustering whatsoever \citep{bor}. An intriguing
result from these wide field surveys is that the high-redshift ULIRGs
appear to trace moderately bright X-ray survey sources on the sky, even
though the direct overlap between the two populations is minimal
\citep{alm}. This could suggest that the high-redshift sub-mm and X-ray
populations are tracing the same, overdense dark matter halos. More recent
efforts to measure SMG clustering using spectroscopic redshifts have met
with greater success, resulting in a reliable detection of clustering for
the first time \citep{bla4}. The strength of the clustering, while
significant, is not particularly strong, at $r_{0}=6.9\pm2.2$
(comoving). This clustering strength is consistent with high-redshift
ULIRGs residing in $\sim10^{13}M_{\odot}$ halos which will eventually
become the cores of rich clusters in the local Universe, but the relatively
high redshift of the sources in this sample, coupled with the significant
error on their clustering and the unknown growth modes of dark matter
overdensities (see e.g. Matarrese et al. 1997; Moscardini et al. 1998)
leaves other possible interpretations open.  More precise measurements of
both high-redshift ULIRGs, and other ULIRGs at lower (and higher) redshifts
are needed to resolve this issue. The new Spitzer ULIRG surveys, sampling
much larger volumes and linear dimensions than current submm/mm surveys,
can be expected open up this field of enquiry dramatically.

\section{Open Questions and New Directions}\label{sec:future}

An incredible amount of energy has been spent trying to understand ULIRGs
in the decade since the Sanders \& Mirabel (1996) review, and this review
has not been able to do adequate justice to all this effort.  To summarize
the current state of knowledge we address the a few of the most pressing
questions from our own perspective.

In the relatively local Universe, ULIRGs are almost all associated with
major mergers of gas rich systems.  Evolutionary schemes whereby ULIRGs
transition into optical QSOs at the end stages of a merger, with an
elliptical as the end result, are highly attractive and well supported by
simulations.  The reality is obviously complex, however, depending strongly
on the details of the merger encounter, in particular bulge mass.
Observationally it is not necessarily the case that all ULIRGs represent
late stages in the merger sequence, however the most luminous systems do
seem to represent the late stages, implying that it requires the
extraordinary high density conditions of the final merger stages to produce
these tremendous levels of power.  Evidence that some optical QSOs may have
more massive hosts than ULIRGs, or inhabit richer environments, would
indicate that the true situation is more complex, but matching
AGN/ULIRG samples for such studies is very challenging.

Merger sequences connect starbursts and AGN tightly together via the
availability of fuel to build the stellar and SMBH masses, and to fuel the
AGN.  AGN orientation schemes also tie ULIRGs and QSOs tightly together
since off-axis QSOs will be obscured and may appear as ULIRGs.  Therefore
the empirical determination of the relative power of AGN and starbursts for
both infrared-selected ULIRGs and optically-selected ULIRG QSOs is a
paramount question. Most ULIRGs are thought to be powered mostly by
starbursts however indicators of AGN activity increase with luminosity,
recognised from near- and mid-infrared spectroscopy, X-ray detections or
VLBI radio imaging.  All of these indicators are tricky however and,
because nuclear (Starburst) and gravitational accretion (AGN) power often
compete for dominance we need more robust estimators of the SFR and AGN
activity to compare with merger stage, morphology, mass, etc. to
disentangle an evolutionary sequence from other effects.  For example,
low/high-resolution mid-IR spectroscopic indicators can be obscured;
Spitzer will do an excellent job of measuring diagnostic features from
5--40$\mu$m, but interpretation will require an understanding of extinction
as a function of wavelength throughout the spectrum as well as modeling
with full radiative transfer treatment.  X-ray columns can be high enough
to severely limit AGN detection; separating soft (starburst or reflected
AGN) vs hard (AGN or X-ray binary) emission will require sensitivity at
higher energies with Con-X and XEUS.  High-resolution radio imaging has to
be capable of mapping at brightness temperatures above $10^6$K to
distinguish jet structures from RSN complexes, because starburst and
radio-quiet AGN emission cannot be distinguished on the basis of power
alone.  But we do not yet even understand the complex radio supernovae in
the one, closest, well studied system, Arp 220, where questions remain
about the nature of the LRSN, the IMF in the luminous starburst, etc.  Such
imaging observations are time consuming, but need to be expanded to much
larger samples.  One possible line of inquiry may be to obtain BH mass
estimates from the molecular gas kinematics with ALMA ($r \sim 20$pc) as
has been done for NGC 4258 in H$_{2}$O.  Other questions remain, {\it eg.}: 
What are the compact masers telling us?  Are they pointers to AGN/mass
concentrations?
   
ULIRGs increase dramatically in importance with redshift, from being a very
minor component at z$\sim$0 to possibly dominating the energy density at
z$>$2, and ULIRG populations at z$>$2 are much more extreme beasts than at
low redshift, with larger IR luminosities, gas masses and, in some cases,
larger ratios of infrared-to-optical luminosity than local ULIRGs, by
factors of 10 to 100. They also show an increase in space density of a
factor of 100-1000 \citep{cha05,lef,perz}, though this number is very
uncertain because the low and high redshift LFs barely overlap in
luminosity range; at low redshift the number density of HLIRGs is
intrinsically extremely small, while at high redshifts, submm surveys are
only sensitive {\it only} to systems above a few ${\times}
10^{12}L_{\odot}$.  Wide area Spitzer luminosity functions can be expected
to bridge this gap (Lonsdale et al. 2006b, Babbedge et al. 2006).  There is
clear evidence that at high redshift many ULIRGs are triggered by mergers
of gas rich systems, but there are also some striking differences from
local systems: extended radio emission and some evidence for very large
cool disk systems, which may not even qualify as starbursts, but rather
star formation ocurring in a more-or-less quiescent fashion in these very
large disks.

As noted above, the incidence of AGN seems to increase with luminosity, and
this is also a strong redshift effect since the most luminous systems are
only known at higher redshifts.  As also noted above, it remains remarkably
difficult to determine the relative power contribution from star formation
vs.  AGN activity in many ULIRGs.  Many lines of argument are used to
conclude that star formation powers most ULIRGs at all redshifts.  These
include the energy reservoir available from nucleosynthesis relative to
SMBH accretion; low AGN power deduced from absorption-corrected hard X-ray
detections; the number of obscured AGN required to explain the XRB and the
correlation between bulge and SMBH masses which suggest bulge growth events
must accompany SMBH growth.  Arguments for a stronger importance for AGN
power include that luminosities above 10$^{13}L_{\odot}$ are very extreme for a
starburst event while many high-redshift QSOs are known with such
luminosities; a large number of QSO-powered ULIRGs is expected in AGN
unification-by-orientation schemes; hard X-ray absorption correction is
difficult, and Compton thick AGN such as Mrk 231 are very difficult
to detect at high redshift.

Perhaps one of the strongest lines of evidence one way or the other comes
from the impressive correlation between warm dense molecular gas traced by
the HCN molecule with infrared luminosity \citep{gs04}, which extends
across 4 orders of magnitude in luminosity from 10$^9$ to 10$^{13}$.
There's little doubt that low luminosity systems are powered by disk star
formation, so this relation provides support for the picture that even
HLIRGs can be powered by star formation.  HCN ULIRG detections are limited at 
high luminosity/redshift, however, and they tend to be displaced above the 
lower luminosity redshift relation towards excess $L_{IR}/M_{HCN}$ 
\citep{carilli}, which is consistent with the presence of an 
an additional, AGN, power source but is also within the scatter of the lower
redshift starburst relation.   Spitzer will be able to provide
deep insights into this debate because the presence of a significant AGN
power source will be evident the mid-infrared spectral shape, as
demonstrated by the recent flurry of papers reporting the discovery of
large populations of highly obscured QSO/ULIRGs
\citep{martinez,poll06,yan,hou,donley}.

The causes for the dramatic differences in ULIRG number density and nature
since z$\sim$3 must lie with the evolving matter density field and galaxy
formation processes.  Since interactions and mergers are clearly implicated
in most ULIRG activity, the most natural assumption to make is that these
objects follow 'the action'.  They are, after all, the most dramatic active
events that occur in the Universe, the prima donnas of the show.  At a
given epoch they are likely to occur where the most dramatic gas rich
mergers are happening, as illustrated in the simulation of K. Nagamine
(private communication) in Figure \ref{lssevo}.  At the earliest epochs of galaxy
formation the action is at the sites of future rich clusters, while at
current times loose groups are ideal future ULIRG sites since there are a
few close galaxies with moderate relative velocity, and gas has not been
exhausted yet.  At intermediate redshifts, the best ULIRG sites may be the
interaction zones in cluster-cluster mergers.  These ideas are consistent
with the sparse currently available observations of ULIRG environments at
various redshifts.

So what are the key unanswered questions and where do we go from here to
answer them?  So far ULIRG samples are quite sparse - numbering in the
hundreds over a huge redshift range.  We clearly need much larger,
volume-limited samples.  Moreover there are serious selection biases since
high-redshift systems have been selected at sub-mm wavelengths by their cool
dust, and substantially warmer systems will have been missed.  Therefore we
need to understand the full range of ULIRG SEDs and obtain complete samples
with well measured SEDs via selection from the mid-IR to the sub-mm.
Spitzer will clearly revolutionize this field, bringing in the warm-samples
to complement the sub-mm ones, but also selecting PAH-dominated objects
as the mid-IR PAH features redshift into the 24$\mu$m band.  
Larger, deeper sub-mm \& mm surveys are also
required to reach down the luminosity function at z$>$1 and combat the
effects of cosmic variance.

To complement the ULIRG samples it's obviously important to study complete
QSO samples in the same volumes of space and to determine the lifetimes of
both types of object, to directly investigate their
evolutionary relationships.  A tricky aspect to this work will be the fact
that ULIRG and QSO luminosities are predicted by the most recent
simulations to vary very dramatically on short timescales (as are 
SMBH and stellar masses), especially as
the SMBH building and AGN accretion phase accelerates, so matching samples
by luminosity or mass is probably going to match apples and oranges.  

Key to understanding the role of ULIRGs in galaxy formation - their connections
to other kinds of systems, their progenitors and their descendents, the
connections between bulge and SMBH building and the relative importance of 
starburst {\it vs} quiescent star formation modes -- is to determine the host 
galaxy properties of ULIRGs and QSOs, especially stellar masses and 
morphologies, and to determine the richness of their environments and their 
clustering properties as a function of redshift.

The future holds great promise for answering these questions. Measuring the
clustering of ULIRGs requires wide area far-IR/sub-mm surveys, which are or
will be forthcoming from Spitzer, SCUBA-2, APEX, ASTRO-F, Herschel and
WISE. These surveys will also provide the large samples of ULIRGs needed to
measure the range in dust temperatures in the distant ULIRG populations,
with detailed followup provided by other groundbased facilities. Measuring
sizes, morphologies, and stellar masses, assessing the impact of
gravitational lensing and determining environmental richness, on the other
hand, require deep pointed observations in the near-IR to radio.  Here the
key facilities are Spitzer, JWST, ALMA, the EVLA and SKA.  Imaging in
molecular and other submm/mm lines will also be possible with ALMA.

\section{Acknowledgements}

We are deeply indebted to Colin Borys, Jason Surace, Kentaro Nagamine,
C. Kevin Xu \& Lee Armus who provided original figures from their recent
work for this review.  The Spitzer IRS ULIRG team who responsible for the
spectra shown and described in Section 3 include B. Thomas Soifer, James
Houck, Lee Armus, Vassilis Charmandaris, Henrik Spoon, Jason Marshall,
Jeronimo Bernard-Salas \& Sarah Higdon.  CJL and HES are indebted to Yu Gao
and the organizers of the Lijiang Workshop on "Extreme Starbursts" which
contributed greatly to the focus of this review.



\end{document}